\documentclass[
aps,
twocolumn,
amsmath,
amssymb,
floatfix,
showpacs,
superscriptaddress,
nofootinbib,
longbibliography
]{revtex4-2}
\usepackage[T1]{fontenc}
\usepackage{mathtools}
\usepackage{braket}
\usepackage{bm}
\usepackage{MnSymbol}
\usepackage{graphicx}
\usepackage{tikz}
\usepackage{pgffor}
\usepackage{blkarray}
\usepackage{sidecap}

\usepackage[dvipsnames]{xcolor}
\usepackage{enumitem}
\usepackage[normalem]{ulem}
\usepackage{silence}

\usepackage[
colorlinks,
linkcolor=NavyBlue,
citecolor=NavyBlue,
urlcolor=NavyBlue
]{hyperref}

\mathchardef\mhyphen="2D

\newcommand\bea{\begin{eqnarray}}
\newcommand\eea{\end{eqnarray}}
\newcommand\beq{\begin{equation}}
\newcommand\eeq{\end{equation}}

\definecolor{lime}{HTML}{A6CE39}

\DeclareRobustCommand{\orcidicon}{%
\hspace{-1.0mm}
\begin{tikzpicture}
  \draw[lime, fill=lime] (0,0)
  circle [radius=0.15]
  node[white] {{\fontfamily{qag}\selectfont \tiny \,ID}};
  \draw[white, fill=white] (-0.0525,0.095)
  circle [radius=0.007];
\end{tikzpicture}
\hspace{-3.0mm}
}

\foreach \x in {A,...,Z}{%
  \expandafter\xdef\csname orcid\x\endcsname{%
    \noexpand\href{https://orcid.org/\csname orcidauthor\x\endcsname}%
    {\noexpand\orcidicon}}%
}

\begin{document}
\title{Emergent topology of flat bands in a twisted bilayer $\alpha-T_3$ lattice}
\author{Gourab Paul$^\clubsuit$\orcidA{}}
\email[]{p.gourab@iitg.ac.in} 
\author{Srijata Lahiri$^\clubsuit$\orcidB{}}
\email[]{srijata.lahiri@iitg.ac.in} 

\author{Kuntal Bhattacharyya\orcidC{}}
\email[]{kuntalphy@iitg.ac.in} 

\author{Saurabh Basu}
\email[]{saurabh@iitg.ac.in}
\affiliation{Department of Physics, Indian Institute of Technology Guwahati, Guwahati-781039, Assam, India}

\def\thefootnote{$\clubsuit$}\footnotetext{These authors contributed equally to this work}

\begin{abstract} 
We provide theoretical evidence of the emergent topology in a twisted bilayer system, despite the corresponding monolayer hosting topologically trivial properties.
To this end, we investigate an interesting interplay of destructive interference due to lattice geometry and band folding due to enlargement of the unit cell in generating and subsequently modifying the band topology in a twisted bilayer $\alpha-T_3$ system. 
The large degeneracy of the emergent flat band in the dice limit of the $\alpha-T_3$ lattice is removed by aligning with h-BN layers, resulting in the formation of sub-bands with varied topological characteristics.
Remarkably, while the sub-band near charge neutrality exhibits a trivial behavior, a topologically non-degenerate singular sub-band emerges away from it.
The topological band remains isolated from the rest of the bands for a substantial area in the $\alpha-\theta$ plane (where $\alpha$ and $\theta$ correspond to the hopping ratio and twist angle, respectively) while exhibiting multiple phase transitions as a function of the aforementioned parameters via hybridization with its nearest bands.
Finally, we study the evolution of the hybrid Wannier charge center and the Chern number to characterize the different emergent topological phases and corroborate our findings.

\end{abstract}

\maketitle
\noindent \textit{Introduction.-} The enormous modification of the properties of monolayer graphene induced by ``stacking and twisting'' of two individual layers of the same, has been under persistent focus in recent years both in the theoretical and experimental front, owing to the multitude of exotic physical phenomena that emerge in the twisted bilayer systems~\cite{Cao2020A, Cao2020b, Cao2018A, Tanaka2025, Ledwith2025A, Ledwith2025B, Khalaf2022, Song2022, Bernevig2021A, Song2021,Bernevig2021B,Lian2021, kolář2025, Perea-Causin2025}.
In the small twist angle regime, the moiré pattern formed by the two layers of graphene leads to an enhancement in the unit cell dimension, causing a considerable band folding.
This, in turn, results in the suppression of kinetic energy near charge neutrality, leading to the formation of flat bands that yield a perfect playground for exploring the interaction effects that conspicuously influence the electronic properties~\cite{Bistritzer2011}.
The vanishing of the Fermi velocity in the flat bands, in fact, leads to numerous emergent phenomena heavily dependent upon the filling factor and aided by strong correlations.
These include the Mott insulating state~\cite{Po2018}, anomalous quantum Hall effect~\cite{Serlin2020}, unconventional superconductivity~\cite{Cao2018B}, and ferromagnetism~\cite{Sharpe2019}, among several others.
\par While a geometric twist, as observed in moiré systems, does give rise to the emergence of flat bands with enhanced interaction effects, similar bands with suppressed kinetic energy can also occur in several other lattice systems with specific sublattice arrangements such that quantum interference causes the wavefunctions to be localized over specific lattice sites, such as the Kagome~\cite{Kang2020, Jiang2019, Taguchi2025} and Lieb lattices~\cite{Mukherjee2015, Jiang2019, Jiang2020, Chen2024}.
Another primary example of such a system includes the dice lattice~\cite{Vidal1998, Wang2011, Mondal2023}, hosting fermionic particles of pseudospin-$1$, which comprises of graphene-like hexagonal unit cells (composed of sublattices \textit{A} and \textit{B}) together with an extra atom at the center of the hexagon, denoted by \textit{C}.
While particles are allowed to hop between the \textit{B} and \textit{C} sublattices, the hopping amplitude between \textit{A} and \textit{C} is quenched.
This leads to the localization of states at the sublattice \textit{B}, resulting in the formation of flat bands.
\begin{figure}[h]
    \centering
    \includegraphics[width=0.95\linewidth]{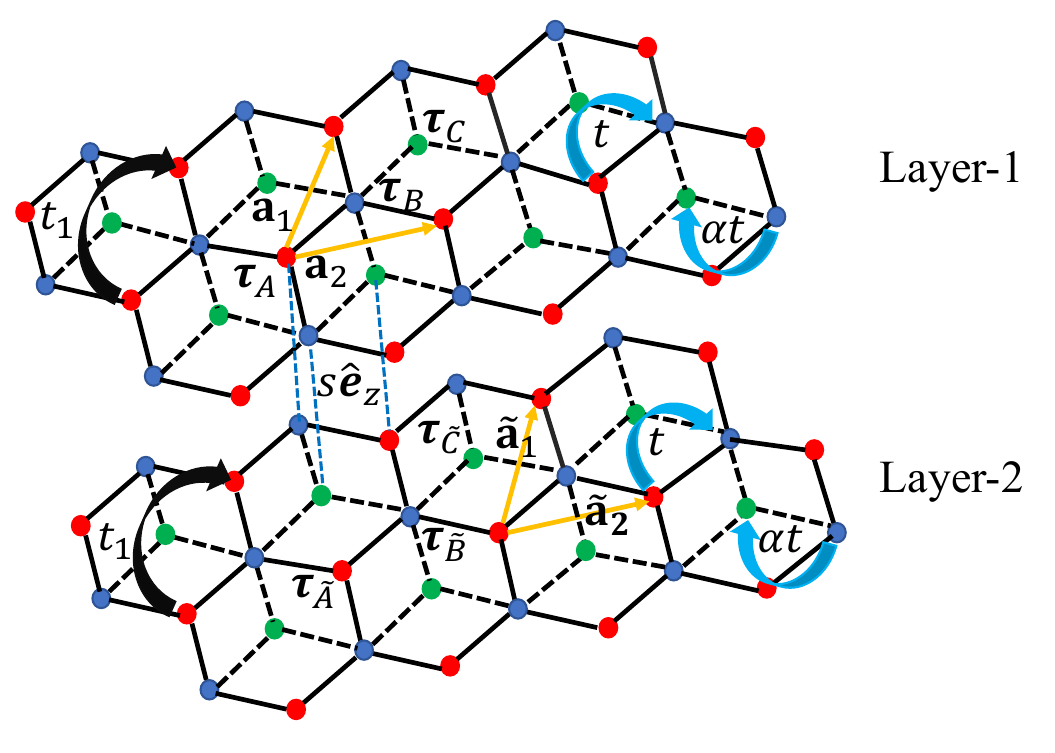}
    \caption{Schematic illustration of the \textit{twisted bilayer} $\alpha-T_3$ lattice with $A$–$B$ (Bernal) stacking configuration. Red, blue, and green dots represent the $A$, $B$, and $C$ sublattice atoms, respectively. The primitive lattice vectors of layer-1 (layer-2) are denoted by $\mathbf{a}_1$, $\mathbf{a}_2$ ($\tilde{\mathbf{a}}_1$, $\tilde{\mathbf{a}}_2$), and the corresponding sublattice position vectors are indicated by $\mathbf{\tau}_X$ ($\mathbf{\tau}_{\tilde{X}}$), with $X = A$, $B$, $C$ and $\tilde{X} = \tilde{A}$, $\tilde{B}$, $\tilde{C}$.
    }
    \label{fig:Figure-1}
\end{figure}
Despite the valence and conduction bands of the \textit{pristine} nearest-neighbor (NN) dice lattice resembling graphene, the triply degenerate bands at the $K$ and $K^{\prime}$ points in the Brillouin zone (BZ) of a dice system garner zero Berry phase, making it non-topological~\cite{Dey2018}.
In recent years, the dice lattice together with its variant, namely the $\alpha-T_3$ lattice, has been rigorously explored in the context of first and second order topological phase~\cite{Lee2024, Dey2019, Kunst2018, Ye2024}, orbital~\cite{Tamang2023}, optical conductivity~\cite{Illes2015}, magnetotransport properties~\cite{Li2022}, stacked systems \cite{Dima1, Dima2}, Floquet drive \cite{Lee2}, transport \cite{FM1, FM2} etc.
The $\alpha-T_3$ lattice provides a generalized framework, smoothly interpolating between the graphene ($\alpha=0$) and the dice ($\alpha=1$) limits, via tuning the ratio $\alpha$ which controls the \textit{B}$\rightarrow$\textit{C} hopping strength $\alpha t$ ($\alpha\in[0, 1]$), whereas the \textit{A}$\rightarrow$\textit{B} hopping is maintained at $t$. \textcolor{black}{The next-nearest-neighbor (NNN) hopping amplitude between the $A$$\rightarrow$$A$, $B$$\rightarrow$$B$, and $C$$\rightarrow$$C$ sublattice atoms is represented by $t_1$.}
\par
Efforts to amalgamate the effects of quantum interference (as present in the dice lattice) with band folding (as present in moiré lattices) have recently been undertaken, with several works exploring the effects of twist in the already prevalent flat bands in bilayer dice systems~\cite{Ma2024, Zhou2024}. 
\begin{figure}
    \centering
    \includegraphics[width=0.9\linewidth]{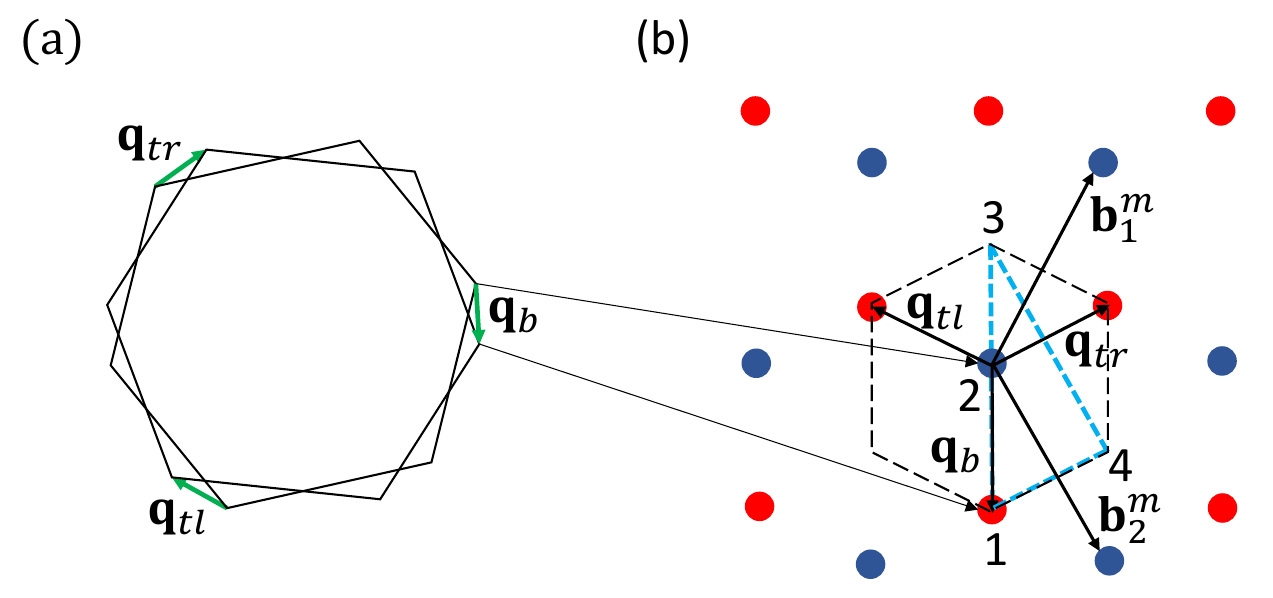}
    \caption{(a) Brillouin zones of the top and bottom layers of the \textit{twisted bilayer} $\alpha-T_3$ lattice, rotated by a relative twist angle $\theta$ with respect to each other. (b) The corresponding MBZ is characterized by the moiré reciprocal lattice vectors $\mathbf{b}_1^m = \dfrac{8\pi \sin(\theta/2)}{\sqrt{3}a} \left( \dfrac{1}{2}, \dfrac{\sqrt{3}}{2} \right)$ and $\mathbf{b}_2^m = \dfrac{8\pi \sin(\theta/2)}{\sqrt{3}a} \left( \dfrac{1}{2}, -\dfrac{\sqrt{3}}{2} \right)$, respectively. \textcolor{black}{The high-symmetry path, shown by the blue dotted line, is chosen as $1 \rightarrow 2 \rightarrow 3 \rightarrow 4 \rightarrow 1$, corresponding to $1 \equiv K^{\prime}$, $2 \equiv K$, $3 \equiv \Gamma$, and $4 \equiv \Gamma$ points of the MBZ.}}
    \label{fig:Figure-2}
\end{figure}
However, despite the twisted bilayer dice lattice being thoroughly explored, a rigorous investigation into the topology of the twisted bilayer $\alpha-T_3$ lattice has been missing from the literature.
Such a system in fact presents a more general playground allowing the behavior of the flat bands to be modulated not only by the twist angle $\theta$, but also by the hopping ratio $\alpha$, where the latter is known to significantly modify the band topology in relevant monolayer lattices \cite{Islam2024, Bhattacharyya2024} owing to the destructive interference mentioned above.
At this point, the questions that we ask are the following: Does the stacking and twisting of the non-topological flat bands already present in the monolayer $\alpha-T_3$ lattice lead to emergent topology in the bilayer system?
If yes, how do the topology as well as flatness of these bands behave under the dual interplay of band folding (controlled by $\theta$) and quantum interference (controlled by $\alpha$).
To answer these questions, we start with a bilayer $\alpha-T_3$ lattice first in the dice limit to obtain a highly degenerate middle band (near charge neutrality) formed due to twisting. 
To lift the degeneracy of this band, we introduce h-BN layers (rendering an $MS_z$ term), thereby resulting in the formation of several isolated sub-bands.
Remarkably, the topological properties of these sub-bands exhibit considerable variation, with the sub-bands near charge neutrality behaving as trivial, while those away from it manifest robust topology.
Furthermore, the non-trivial bands undergo several phase transitions (gap closing) as a function of $\alpha$ and $\theta$, showing the existence of smoothly tunable topology mediated by band folding, as well as quantum interference. We establish the topological nature of these phase transitions using the Chern number together with the evolution of the hybrid Wannier charge centers that provide robust support to our findings.
The bandwidth of the topological band is also observed to be majorly affected by both $\alpha$ and $\theta$, indicating the influence of both interference and band folding in suppressing the kinetic energy associated with the topological flat band.\\
\textit{Model Hamiltonian.-}
We begin by considering a bilayer $\alpha-T_3$ lattice (Fig.~\ref{fig:Figure-1}), where the top layer (layer-1) is rotated by an angle $\theta/2$ whereas the bottom layer (layer-2) is given a rotation $-\theta/2$. The corresponding continuum model for the \textit{twisted bilayer} $\alpha-T_3$ system can be constructed directly following the Bistritzer–MacDonald formalism~\cite{Bistritzer2011} as;
\begin{equation}
    H(\theta) = 
\begin{pmatrix}
H_0\left( \frac{\theta}{2} \right) & U \\
U^\dagger & H_0\left( -\frac{\theta}{2} \right)
\end{pmatrix},
\label{Twisted_Bilayer_Ham}
\end{equation}
where $H_0$ corresponds to the intralayer Hamiltonian of the individual $\alpha-T_3$ layers and $U$ denotes the interlayer tunneling matrix.
It is to be noted that the Hamiltonian for a monolayer $\alpha-T_3$ lattice when twisted by an angle $\theta$ [denoted by $H_0(\theta)$], \textcolor{black}{and expanded around the rotated valley point $\mathbf{K}_\theta$ with the inclusion of NNN hopping, can be written as;
\begin{equation}
    H_{\textbf{K}_\theta}(\textbf{q}_\theta) = v_F q_{\theta}
\begin{pmatrix}
0 & e^{-i(\theta_{\textbf{q}} - \theta)} & 0\\
e^{i(\theta_{\textbf{q}} - \theta)} & 0 & \alpha e^{-i(\theta_{\textbf{q}} - \theta)} \\
0 & \alpha e^{i(\theta_{\textbf{q}} - \theta)} & 0
\end{pmatrix} + t_2 q_{\theta}^2~\mathcal{I}_3,
\end{equation}}
\textcolor{black}{where $t_2$ = $\frac{\sqrt{3} a t_1}{2 t} v_F$, $\mathcal{I}_3$ =  $\begin{pmatrix}
1 & 0 & 0\\
0 & 1 & 0 \\
0 & 0 & 1
\end{pmatrix}$, and $\mathbf{q}_\theta = q_\theta \left(\cos\theta_{\mathbf{q}}, \sin\theta_{\mathbf{q}}\right)$ denotes the momentum measured with respect to $\mathbf{K}_\theta$, the rotated $K$ point in the Brillouin zone of the twisted monolayer $\alpha-T_3$ lattice.} In the remainder of the text, we denote the Dirac points of layer-$1$ (layer-$2$) by $K$ ($\tilde{K}$) and $K'$ ($\tilde{K'}$). Extensive calculations pertaining to the monolayer $\alpha-T_3$ lattice have been provided in the Supplemental Material (SM).
\par
\begin{figure}
    \centering
    \includegraphics[width=1.0\linewidth]{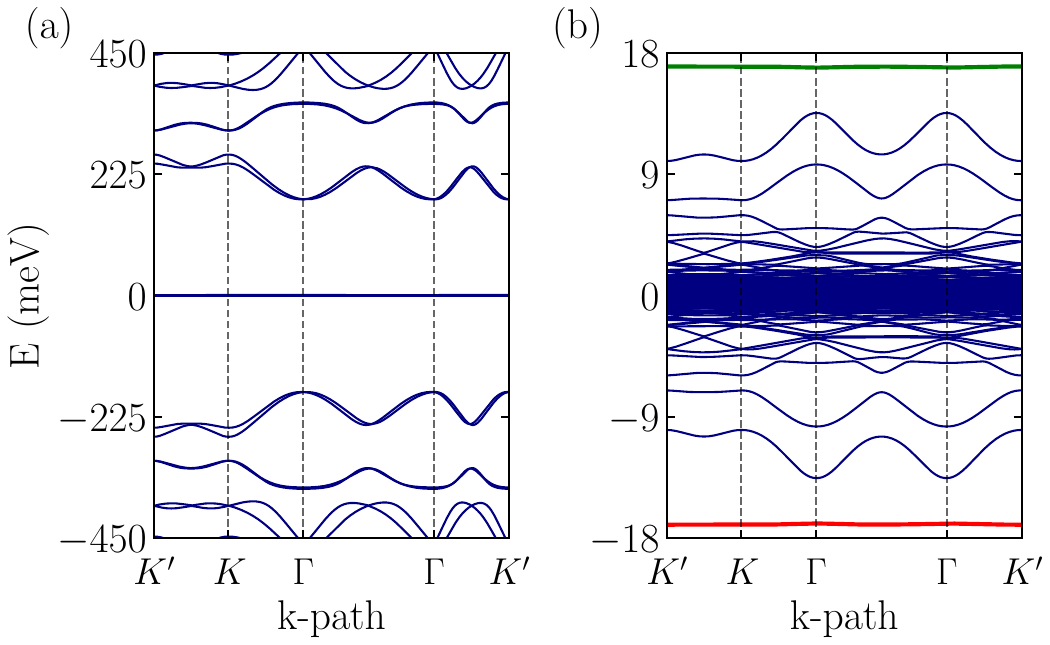}
    \caption{\textcolor{black}{Band structure of the \textit{twisted bilayer} $\alpha-T_3$ lattice in the dice ($\alpha=1$) limit, where panel (a) corresponds to the energy eigenvalues evaluated at the magic angle $\theta = 1.08^\circ$ in the chiral limit ($w_1 = 0$, $w_2 = 110.7$~meV, $w_3 = 0$), with the Fermi velocity $v_F = 6326.1$~meV$\cdot$\AA~and $t_2 = 0.001~v_F$. The panel (b) is obtained with an alignment of two h-BN substrate layers at $A$ and $C$ sublattice sites, where one layer is placed above the top $\alpha-T_3$ layer and the other beneath the bottom $\alpha-T_3$ layer.
    }}
    \label{fig:Figure-3}
\end{figure}

 Now we consider the interlayer Hamiltonian $U$, with the NN hopping from sublattice site $X$ ($X \in A, B, C$) in layer-1 to the nearest sublattice site $\tilde{X}$ ($\tilde{X} \in \tilde{A}, \tilde{B}, \tilde{C}$) in layer-2. For a small twist angle $\theta$, we only consider the scattering processes to occur between the $K$
and the $\tilde{K}$ valley, 
neglecting the intervalley scattering between the two layers, thus ensuring a conservation of the valley degrees of freedom.
 It is important to mention here that the presence of this valley $\mathcal U$($1$) symmetry in the \textit{twisted bilayer} $\alpha-T_3$ lattice ensures a similar treatment for the $K'$ valley as well.
 However, for the purpose of our study, we now focus mainly on the $K$ valley.
 In the vicinity of $K$, the matrix elements of the interlayer Hamiltonian are given by~\cite{Koshino2015, Ma2024};
\begin{eqnarray}
    U_{\tilde{X},X}(\mathbf{q}, \tilde{\mathbf{q}}) = \sum_{\mathbf{G}, \tilde{\mathbf{G}}} w_{\tilde{X},X}(\mathbf{q} + \mathbf{K} + \mathbf{G}) \, e^{-i \mathbf{G} \cdot \boldsymbol{\tau}_X + i \tilde{\mathbf{G}} \cdot \boldsymbol{\tau}_{\tilde{X}}} \nonumber \\ 
    \times \quad \delta_{\mathbf{q} + \mathbf{K} + \mathbf{G}, \tilde{\mathbf{q}} + \tilde{\mathbf{K}} + \tilde{\mathbf{G}}}.
\end{eqnarray}
Here, $\mathbf{G} = m_1 \mathbf{b}_1 + m_2 \mathbf{b}_2$ and $\tilde{\mathbf{G}} = m_1 \tilde{\mathbf{b}}_1 + m_2 \tilde{\mathbf{b}}_2$, where $\mathbf{b}_1$ ($\tilde{\mathbf{b}}_1$) and $\mathbf{b}_2$ ($\tilde{\mathbf{b}}_2$) are the reciprocal lattice vectors of layer-1 (layer-2), corresponding to its primitive lattice vectors  $\mathbf{a}_1$ ($\tilde{\mathbf{a}}_1$) and $\mathbf{a}_2$ ($\tilde{\mathbf{a}}_2$) respectively.
For the long-period moiré superlattice, the reciprocal lattice vectors of the two layers are related by a rotation matrix $\mathcal{R}_{\theta}$, such that $\tilde{\mathbf{b}}_i = \mathcal{R}_{\theta} \mathbf{b}_i$~\cite{Koshino2015}. The vectors
$\boldsymbol{\tau}_{X}$ and $\boldsymbol{\tau}_{\tilde{X}}$ represent the sublattice position of the atoms in layer-1 and layer-2, respectively. Now, we consider $A$-$B$ (Bernal) stacking configuration, under which the sublattice position vector of the atoms in both layers are given as, $\mathbf{\tau}_A = -\frac{1}{3}\left( \mathbf{a}_1 + \mathbf{a}_2\right)$, $\mathbf{\tau}_B = 0$, $\mathbf{\tau}_C = \frac{1}{3}\left( \mathbf{a}_1 + \mathbf{a}_2\right)$, $\mathbf{\tau}_{\tilde{A}} = \mathbf{\tau}_0 + s \hat{e}_z + \frac{1}{3} \left( \tilde{\mathbf{a}}_1 + \tilde{\mathbf{a}}_2\right)$, $\mathbf{\tau}_{\tilde{B}} = \mathbf{\tau}_0 + s \hat{e}_z - \frac{1}{3} \left( \tilde{\mathbf{a}}_1 + \tilde{\mathbf{a}}_2\right)$, and $\mathbf{\tau}_{\tilde{C}} = \mathbf{\tau}_0 + s \hat{e}_z$. Here, $s$ denotes the interlayer separation, and $\mathbf{\tau}_0$ is the relative in-plane translation vector of layer-2 with respect to layer-1. For simplicity, we consider $\mathbf{\tau}_0 = 0$ in our analysis. Incorporating all the considerations above, the interlayer Hamiltonian takes the following form,
\begin{equation}
    U(\mathbf{q}, \tilde{\mathbf{q}}) = U_{\mathbf{q}_b} \delta_{\mathbf{q}-\tilde{\mathbf{q}}- \mathbf{q}_b} + U_{\mathbf{q}_{tr}} \delta_{\mathbf{q}-\tilde{\mathbf{q}}- \mathbf{q}_{tr}} + U_{\mathbf{q}_{tl}} \delta_{\mathbf{q}-\tilde{\mathbf{q}}- \mathbf{q}_{tl}},\label{Interlayer_ham}
\end{equation}
where, the momentum transfer vectors connecting the nearest Dirac points of the two layers in the moiré Brillouin zone (MBZ) (Fig.~\ref{fig:Figure-2}) are given by, $\textbf{q}_b = \frac{8\pi \sin(\theta/2)}{3a} \left( 0, -1\right)$, $\textbf{q}_{tr} = \frac{8\pi \sin(\theta/2)}{3a} \left( \frac{\sqrt{3}}{2}, \frac{1}{2}\right)$, $\textbf{q}_{tl} = \frac{8\pi \sin(\theta/2)}{3a} \left( \frac{-\sqrt{3}}{2}, \frac{1}{2}\right)$. The matrices $U_{\mathbf{q}_b}$, $U_{\mathbf{q}_{tr}}$, $U_{\mathbf{q}_{tl}}$, which describe the interlayer coupling associated with each momentum transfer, take the following forms (see SM):
\begin{align}
U_{\mathbf{q}_b} &=  \begin{pmatrix}
w_1 & w_2 & w_3\\
w_2 & w_1 & w_2 \\
w_3 & w_2 & w_1
\end{pmatrix},\nonumber\\
U_{\mathbf{q}_{tr}} &=  \begin{pmatrix}
w_1 e^{i\phi}& w_2 & w_3e^{-i\phi}\\
w_2e^{-i\phi} & w_1 e^{i\phi} & w_2 \\
w_3 & w_2 e^{-i\phi}& w_1e^{i\phi}
\end{pmatrix},\nonumber\\
U_{\mathbf{q}_{tl}} &= \begin{pmatrix}
w_1 e^{-i\phi}& w_2 & w_3e^{i\phi}\\
w_2e^{i\phi} & w_1 e^{-i\phi} & w_2 \\
w_3 & w_2 e^{i\phi}& w_1e^{-i\phi}
\end{pmatrix}. 
\end{align}
Here, $\phi = 2\pi/3$, and we choose the NN bond length as $d = 1.42 \,\text{\AA}$, same as the lattice constant in twisted bilayer graphene (TBG)~\cite{Ma2024, LopesdosSantos2007}.\\
\textit{Making moiré dice lattice topological.-}
The origin and the topological nature of the flat bands become more evident in the chiral limit~\cite{Tarnopolsky2019}. In our case, for \textit{twisted bilayer} $\alpha-T_3$ lattice, the generalized chiral symmetry is implemented by setting $w_1 = w_3 = 0$, which effectively suppresses interlayer tunneling between identical sublattice sites across the two layers, as well as the tunneling involving $A$ and $C$ sublattices of different layers. As a result, the interlayer Hamiltonian is entirely governed by the parameter $w_2$, which we fix at  $w_2=110.7$ \text{meV}, consistent with the value used in TBG~\cite{Koshino2015, Ma2024}. \textcolor{black}{For the Bloch-band-based effective theory to be valid, the presence of a valley structure is essential. However, the existence of a global flat band in the single-layer dice lattice can undermine this requirement. This issue can be resolved by introducing a NNN hopping $t_1$ into the lattice model, which makes the global flat bands dispersive and renders a valley structure.} Now diagonalizing the Hamiltonian of Eq.\eqref{Twisted_Bilayer_Ham} in the plane-wave basis, we obtain the band structure along the dotted path shown in Fig.~\ref{fig:Figure-2}(b). \textcolor{black}{The vanishing of $w_1$ and $w_3$ with the NNN hopping $t_1$; leads to the emergence of an exactly flat, isolated band at $E = 0$ with substantial degeneracies in the MBZ at magic angle ($\theta = 1.08^\circ$) and in the dice limit ($\alpha = 1$) [Fig.~\ref{fig:Figure-3}(a)].} Due to the pronounced degeneracy of this isolated flat band, any topological characterization becomes ill-defined~\cite{Zhou2024, Ma2024}.
 We refer to this band as the middle band in the following discussions. 
\begin{figure}
\centering
\includegraphics[width=\columnwidth]{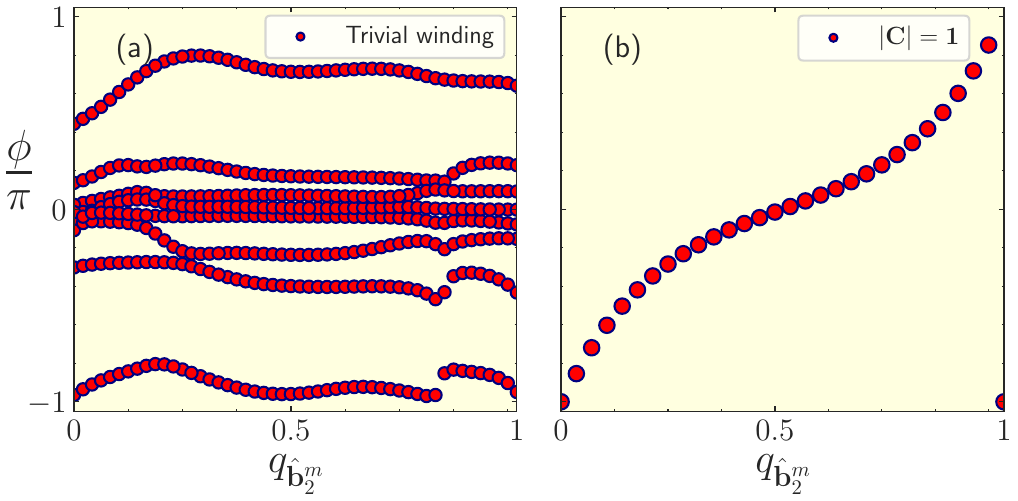}
\caption{Evolution of the hybrid WCC calculated along the direction $\mathbf{\hat{a}}_1^m$ as a function of momenta along the direction $\mathbf{\hat b}_2^m$ in the presence of the h-BN layers for (a) the sub-bands near charge neutrality and (b) the NFB away from charge neutrality.
} \label{fig:Figure-4}
\end{figure}
\begin{figure*}
    \centering
    \includegraphics[width=1.0\textwidth]{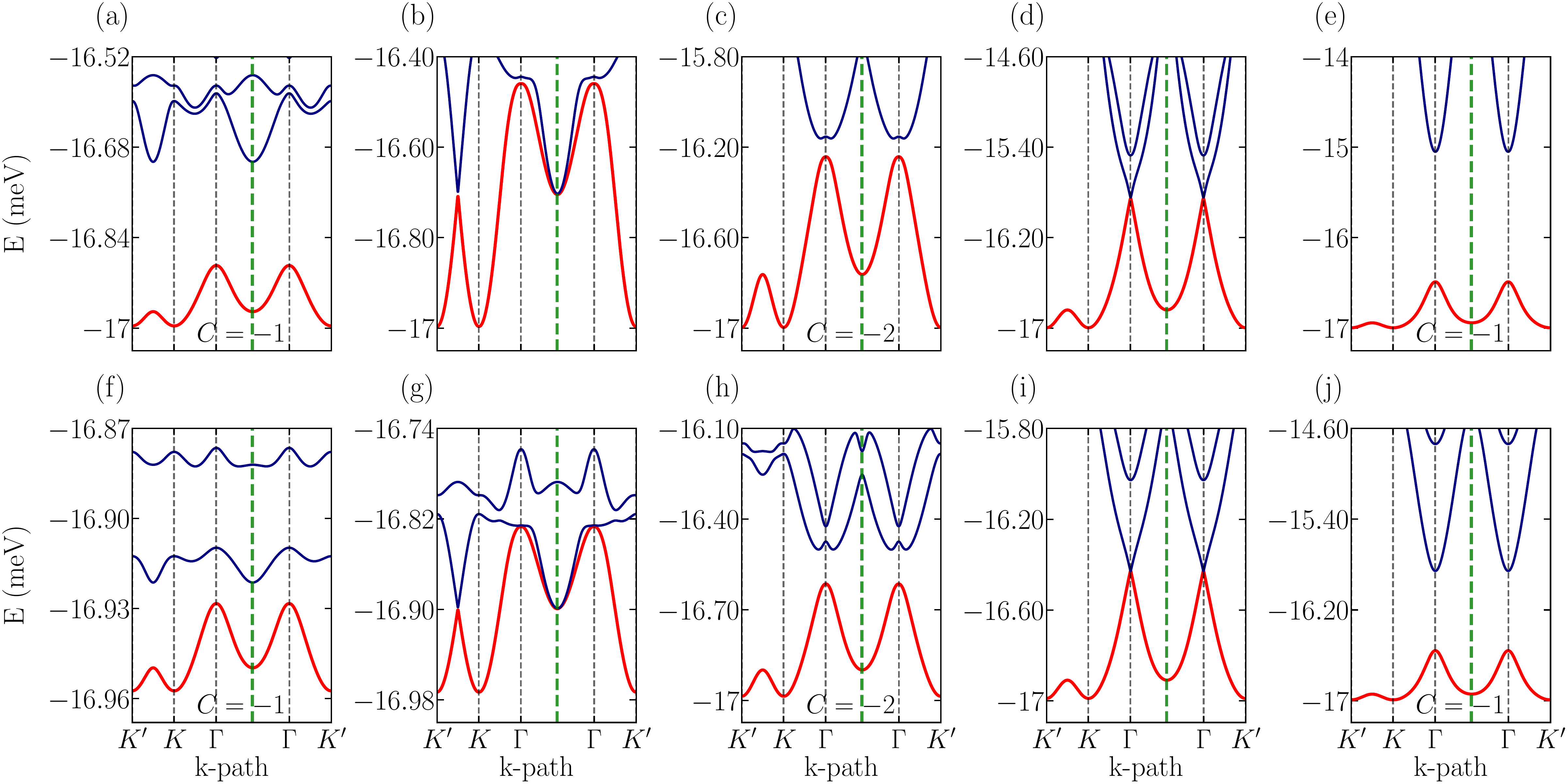}
    \caption{\textcolor{black}{The band structures of the NFB in the dice limit are shown in panels (a)–(e) for increasing twist angles $\theta = 0.32^{\circ}, 0.40^{\circ}, 0.50^{\circ}, 0.62^{\circ},$ and $0.80^{\circ}$, with $t_2 = 0.001~v_F$. Panels (f)–(j) display the evolution of the NFB at fixed magic angle $\theta = 1.08^{\circ}$ for increasing values of $\alpha = 0.30, 0.368, 0.50, 0.567,$ and $0.70$, respectively, with $t_2 = 0.001~v_F$. All other parameters are kept identical to those in Fig.~\ref{fig:Figure-3}(b). The green dashed line, drawn perpendicular to the k-path between two $\Gamma$ points, marks the $M$ point along the high-symmetry path in the MBZ.}}
    \label{fig:Figure-5}
\end{figure*}
\par
With the objective of lifting the degeneracy in the middle band, we now consider the \textit{twisted bilayer} $\alpha-T_3$ lattice to be aligned with two h-BN layer substrates above and below the layer-1 and layer-2, respectively.
We further assume that this alignment is configured such that each h-BN layer couples selectively to the $A$ and $C$ sublattice atoms in the adjacent layer of the $\alpha-T_3$ lattice, while the $B$ sublattice remains entirely unaffected by the substrate potential.
We also consider that the twist angle $\theta_{\text{h-BN}}$ between the top (bottom) h-BN layer and the adjacent $\alpha-T_3$ lattice is close to zero.
Additionally, we presume that the effect caused due to misalignment of the top (bottom) h-BN layer with the bottom (top) $\alpha-T_3$ lattice and h-BN layer is negligible and therefore ignored in our analysis.
This aligned top (bottom) h-BN layer induces two primary effects on the top (bottom) $\alpha-T_3$ lattice.
First, it includes a staggered sublattice potential on the $A$ and $C$ sites of each layer.
This effect is modeled as a staggered ``mass" term $M$ that enters the intralayer Hamiltonian in the $A$ and $C$ sublattice sector, represented by $MS_z$.
We choose $M \approx 17$ meV for our analysis, consistent with the value reported in the studies of TBG~\cite{Zhang2019A,Cea2020,Liu2021,Sun2021}.
Second, there is a moiré potential emerging from the lattice mismatch between the h-BN layer and $\alpha-T_3$ lattice. In line with the treatment adopted for TBG~\cite{Cea2020,Zhang2019A, Zhang2019B}, we neglect this term in our analysis.
It is observed that this alignment causes the degeneracy of the middle band to be considerably lifted, thereby leading to substantial broadening.
Under the combined conditions of chiral limit ($w_1 = w_3 = 0$), magic angle ($\theta = 1.08^\circ$) and dice limit ($\alpha = 1$), it is observed that the system hosts two isolated, non-degenerate nearly flat sub-bands (which were earlier part of the degenerate middle band) at the edges of the broadened spectrum of the middle band.
We then call these isolated sub-bands the nearly flat bands (NFBs) for brevity.
\textcolor{black}{These bands are positioned about the zero energy, located at $E = \pm M$, shown in red and green [Fig.~\ref{fig:Figure-3}(b)].}
However, in the close vicinity of $E=0$, the bands still show considerable overlap, preventing them from being treated individually.
As we move away from the magic angle and the dice limit, the flatness of these bands is lifted, leading to a dispersive band structure across the 
$\alpha-\theta$ plane.
In the subsequent section, we discuss the topological properties of the broad middle band spectrum, primarily focusing on the NFB located at $E = -M$, as a function of $\alpha$ and $\theta$, which are two of the most crucial parameters influencing our system.\\
\textit{Topological characterization of NFBs.-}
To ascertain the topological characteristics of the different regions of the middle band, we first observe the evolution of the hybrid Wannier charge center (WCC) corresponding to the central region of the band~\cite{Taherinejad2014, Soluyanov2011, Lahiri2024}. 
We evaluate the hybrid WCC along the moiré lattice vector $\mathbf{a}_1^m$ while studying its evolution as a function of the momentum vector along the reciprocal lattice vector $\mathbf{b}_2^m$. 
We observe that the evolution of the multi-band hybrid WCC, calculated for a bundle of eight bands near charge neutrality, shows a trivial evolution as a function of $q_{\hat{\mathbf{b}}_2^m}$ [Fig.~\ref{fig:Figure-4}(a)]. 
This indicates that in the vicinity of the band center, the middle band hosts states that are topologically trivial.
Interestingly, however, the single band hybrid WCC calculated for the NFB at the band edge, shows a non-trivial evolution as a function of $q_{\hat{\mathbf{b}}_2^m}$, indicating a Chern number $|C|=1$ [Fig.~\ref{fig:Figure-4}(b)].
It is to be noted here that the flat band of monolayer $\alpha-T_3$ does not show any topological behavior in the dice limit, thereby confirming that the emergent topology in the bilayer system occurs due to band folding induced by twist.
This is one of the key observations of our work.
Having established the topological nature of the non-degenerate NFB, we now focus on its evolution in the $\alpha-\theta$ plane.
The evolution of the NFB extracted using h-BN layers is accompanied by a series of band gap closings and reopenings as a function of $\alpha$ and $\theta$, indicative of probable topological phase transitions that arise from the intricate interplay between the aforementioned parameters.
In the rest of our work, we systematically study the topological evolution of this NFB using a single band Chern number ($C$) evaluated employing Fukui's method~\cite{Fukui2005}. 
For the sake of concreteness, we primarily focus on two specific limits, which are of prime importance for our study, namely the dice limit and the magic-angle limit.
\par First, we analyze the electronic structure of the non-degenerate NFB in the dice limit (fixing $\alpha = 1$) and for a substantial range of the twist angle $\theta$.
It is important to note that away from the magic angle, the NFB no longer remains flat and becomes dispersive.
The corresponding band structures for increasing $\theta$ values ($0.32^\circ$, $0.40^\circ$, $0.50^\circ$, $0.62^\circ$, and $0.80^\circ$) are shown in Figs.~[\ref{fig:Figure-5}(a)-\ref{fig:Figure-5}(e)]. 
At $\theta = 0.32^\circ$, the NFB lies in the topological phase characterized by the Chern number, $C = -1$ [Fig.~\ref{fig:Figure-5}(a)]. 
As we increase the twist angle $\theta$, this NFB undergoes a first order topological phase transition from $C=-1$ at $\theta=0.32^\circ$ to $C=-2$ at $\theta=0.50^\circ$ [Fig.~\ref{fig:Figure-5}(c)],
via a gap closing transition at $\theta=0.40^\circ$ [Fig.~\ref{fig:Figure-5}(b)]. 
It is to be noted here that the gap opening occurs at the $M$ point of the MBZ.
\begin{figure}
    \centering    \includegraphics[width=1\linewidth]{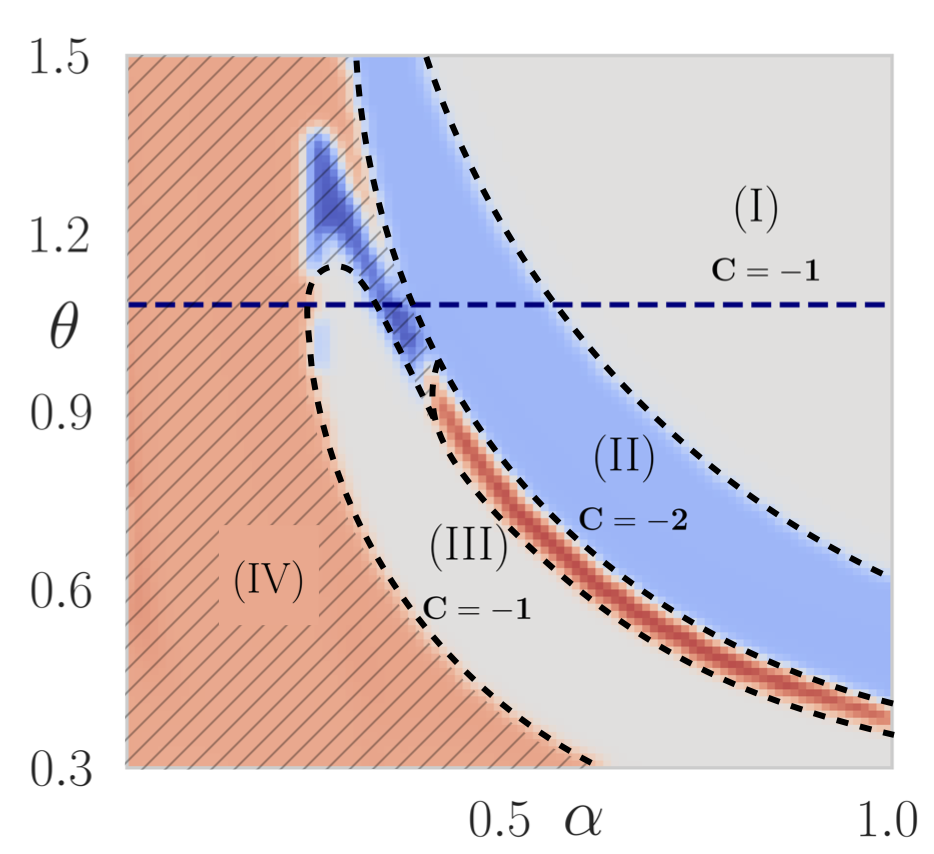}
    \caption{
    The phase plot of the Chern number for the NFB in the $\alpha-\theta$ plane. The horizontal dotted line shows the magic angle limit. The colors separate the areas with varying Chern numbers.
    \textcolor{black}{The regions marked (I), (II) and (III) correspond to topological phases with Chern number $C=-1$, $-2$ and $-1$ respectively. The phase marked (IV) (the hatched area) corresponds to an area of negligible ($10^{-4}$ meV) or no band gap, resulting in an ill-defined topological characterization. The sliver of area separating region (II) and (III) is also a semi-metallic phase with a negligible band gap and undefined topological characterization.}}
    \label{fig:Figure-12}
\end{figure}
On further increasing $\theta$, the system undergoes another gap closing event occurring at $\theta = 0.62^\circ$ [Fig.~\ref{fig:Figure-5}(d)], which leads to another first order topological phase transition of the NFB from $C=-2$ back to $C=-1$ [Fig.~\ref{fig:Figure-5}(e)] at the $\Gamma$ point of the MBZ. 
Beyond $\theta=0.62^\circ$ (up to a certain maximum value considered in the Chern phase diagram shown later), the Chern number associated with the NFB remains unchanged. 
In particular, at $\theta = 1.08^\circ$ (in the dice limit), the band resides in the $C = -1$ topological phase.

\par Now we study the electronic structure of the NFB at the magic angle, fixing $\theta = 1.08^\circ$, simultaneously along with exploring the entire range of $\alpha$ between $0$ and $1$.
Similar to the earlier treatment, the NFB becomes dispersive in nature as we move away from the dice limit, keeping $\theta$ fixed at the magic angle. 
\textcolor{black}{The corresponding band structures for increasing values of $\alpha$ (0.30, 0.368, 0.50, 0.567, and 0.70) are shown in Fig.~[\ref{fig:Figure-5}(f)-\ref{fig:Figure-5}(j)]. 
At $\alpha = 0.30$, the NFB lies in the $C = -1$ Chern phase [Fig.~\ref{fig:Figure-5}(f)].
Now, as we increase $\alpha$, this NFB shows a gap closing transition at $\alpha=0.368$ [Fig.~\ref{fig:Figure-5}(g)], which leads to a first order topological phase transition at the $M$ point of the MBZ from $C=-1$ to $C=-2$ [Fig.~\ref{fig:Figure-5}(h)].
Upon further increasing $\alpha$, this NFB undergoes another gap closing transition occurring at $\alpha = 0.567$ [Fig.~\ref{fig:Figure-5}(i)] at the $\Gamma$ point of the MBZ, leading to a subsequent first order topological phase transition that returns the system to the $C = -1$ phase at $\alpha = 0.70$ [Fig.~\ref{fig:Figure-5}(j)].}
Furthermore, the system remains in this $C=-1$ phase for the larger values of $\alpha$ till the dice limit ($\alpha=1$).
Thus, we establish the emergence of topology at the edge of the middle band of a bilayer $\alpha-T_3$ induced by twist and subsequent phase transitions in the topological band maneuvered by the interplay of the twist angle $\theta$ and the hopping ratio $\alpha$.
The phase plot of the Chern number shown in Fig. \ref{fig:Figure-12}, exhibits a complete variation of the Chern number in the full $\alpha-\theta$ plane. 
The separate areas of varying topology are marked distinctly as (I), (II), and (III) with their respective Chern numbers as $-1$, $-2$, and $-1$. 
Region (IV), comprising low $\theta$ and $\alpha$ values, corresponds to an area of negligible band gap ($10^{-4}$ meV) and therefore an ill-defined Chern number.
Furthermore, the small sliver separating region (II) and (III) corresponds to a semi-metallic phase with a negligible ($\approx 0$ meV) band gap. The edge manifestations of the various topological phases depicted in Fig.~\ref{fig:Figure-5}, have been elaborately discussed in the SM.
 \begin{figure}[h]
    \centering
    \includegraphics[width=1.0\linewidth]{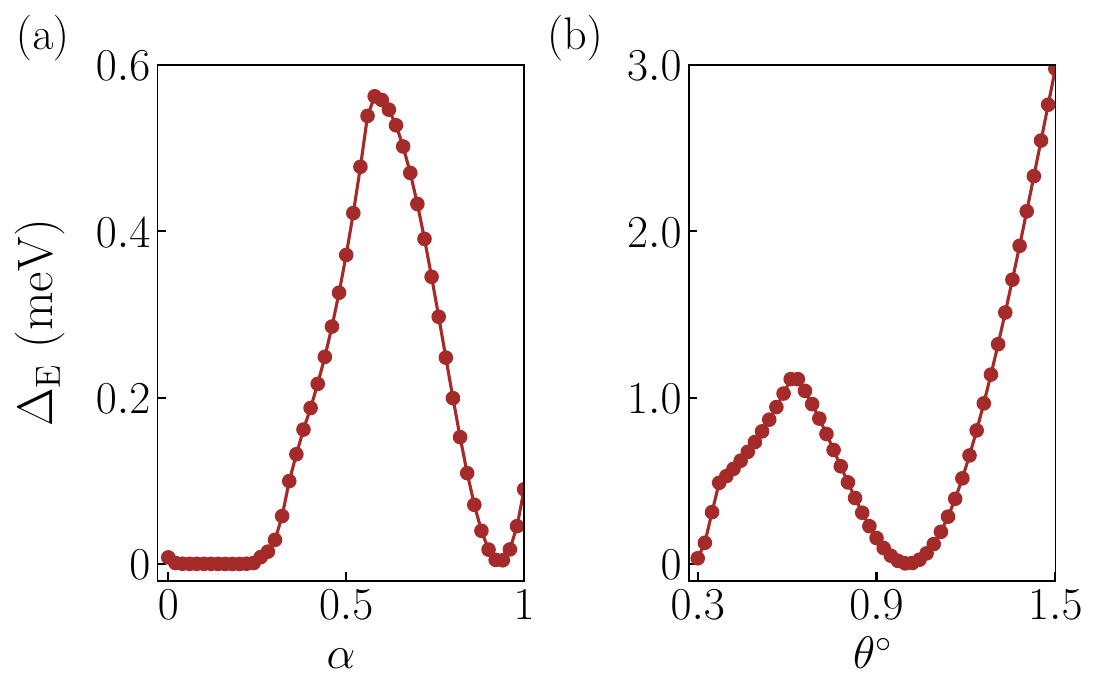}
    \caption{The bandwidth of the NFB is shown (a) as a function of $\alpha$ for $\theta=1.08^\circ$ and (b) as a function of $\theta$ for $\alpha=1$. It is observed that the bandwidth of the NFB increases when deviating from the magic angle limit as well as the dice limit.}
    \label{fig:Figure-6}
\end{figure}
\par Another important aspect of our study is the variation of the bandwidth ($\Delta_E=E^{\text{max}}_{\text{NFB}}- E^{\text{min}}_{\text{NFB}}$) of the NFB at the middle band edge as a function of $\alpha$ and $\theta$.
 The bandwidth of the NFB is shown in Fig. \ref{fig:Figure-6}, first as a function of $\alpha$ in the magic angle limit and then as a function of $\theta$ in the dice limit.
 It is observed that for very low values of $\alpha$ in the magic angle limit, the band is extremely flat ($E^{\text{max}}_{\text{NFB}}\approx E^{\text{min}}_{\text{NFB}}$). 
Interestingly, for intermediate values of $\alpha$, the band flatness is reduced (becomes more dispersive)  despite the system being at the magic angle, with the flatness increasing substantially again for $\alpha$ nearing the dice limit.
In the dice limit, the variation of the bandwidth with $\theta$ shows a predictable trend with the bands being minimally dispersive only around $\theta=1.08^\circ$.
It is therefore evident that the flatness of the topological NFB is substantially influenced not only by the twist angle $\theta$ but also by the hopping ratio $\alpha$, with the band becoming substantially dispersive for the intermediate values of $\alpha$ even at the magic angle limit.
However, it is important to mention here that the value of the bandwidth is visibly low for the entire $\alpha$ range in the magic angle limit, as compared to the variation with respect to $\theta$ in the dice limit.
This establishes that the twist angle has a more pronounced effect on the band flatness when compared to the hopping ratio $\alpha$.\\
\noindent \textit{Conclusion.-}
We observe the emergence of non-degenerate topological sub-bands at the edge of the highly degenerate flat band of twisted bilayer $\alpha-T_3$ lattice in the dice limit and at the magic angle, when the degeneracy is lifted by aligning the individual $\alpha-T_3$ layers with h-BN.
The topology persists in the system as we move away from both $\alpha=1$ and $\theta=1.08^\circ$ limits with the NFB undergoing several gap closing transitions as functions of $\alpha$ and $\theta$.
We probe the topology of these transitions by evaluating the Chern number of the NFB.
Interestingly, the sub-bands near charge neutrality show no topological feature, as is confirmed by the trivial evolution of the multi-band WCC.
The phase diagram plotted in the $\alpha-\theta$ plane shows how the Chern number of the NFB evolves as a function of these parameters, exhibiting several regions of distinct topology characterized by the respective value of their invariants.
Our results are furthermore converged with respect to the number of lattice points taken in the reciprocal space to ensure the authenticity of the results.\\
\textcolor{black}{\textit{Experimental realization.-} It is well known that the dice lattice ($\alpha=1$) can be experimentally realized by growing trilayers of cubic lattices, namely SrTr$\text{O}_3$/SrIr$\text{O}_3$/SrTr$\text{O}_3$, in the $[111]$ direction \cite{Wang2011}. Such heterostructures can generate the low energy electronic states of a dice lattice and thereby provide a platform to materialize our results in the experimental scenario. Furthermore, other experimental platforms where such intricate hopping/tunneling processes can be accurately realized include optical lattices, which allow a robust control over all the system parameters \cite{Dice_exp2, Dice_exp3}. In fact, the realization of the unique hopping structure of dice ($t_\text{AB}=t_\text{BC}$ and $t_\text{AC}=0$) can be more easily realized in optical lattices, even though it may be difficult in real materials. It should also be mentioned here that $\text{Hg}_{1-x}\text{Cd}_x\text{Te}$ with a critical doping can be mapped on to an $\alpha-T_3$ lattice with $\alpha=\frac{1}{\sqrt3}$ \cite{PhysRevB.92.035118}, hence providing another alternative to the formation of the twisted structure at a particular value of $\alpha$.}\\
\textit{Acknowledgments.-}
G.P. acknowledges financial support from the Ministry of Education (MoE), Government of India, through the research fellowship. S.L. acknowledges financial support from the MoE, Govt. of India, through the Prime Minister’s Research Fellowship (PMRF) scheme in May 2022. K.B. acknowledges the Anusandhan National Research Foundation (ANRF), Govt. of India, for providing financial support through the National Post Doctoral Fellowship (NPDF) (File No. PDF/2023/000161).

\bibliography{bibfile}

\clearpage
\onecolumngrid
\widetext


\setcounter{equation}{0}
\setcounter{figure}{0}
\setcounter{table}{0}

\renewcommand{\theequation}{S\arabic{equation}}
\renewcommand{\thefigure}{S\arabic{figure}}
\renewcommand{\thetable}{S\arabic{table}}
\renewcommand{\bibnumfmt}[1]{[S#1]}
\renewcommand{\citenumfont}[1]{S#1}

\newcommand{\bk}{\boldsymbol\kappa}
\newcommand{\SI}{Supplementary Material}

\newcommand{\beginsupplement}{%
  \setcounter{equation}{0}
  \renewcommand{\theequation}{S\arabic{equation}}%
  \setcounter{table}{0}
  \renewcommand{\thetable}{S\arabic{table}}%
  \setcounter{figure}{0}
  \renewcommand{\thefigure}{S\arabic{figure}}%
  \setcounter{section}{0}
  \renewcommand{\thesection}{S\Roman{section}}%
  \setcounter{subsection}{0}
  \renewcommand{\thesubsection}{S\Roman{section}.\Alph{subsection}}%
}

\beginsupplement

\begin{center}
    {\large \textbf{Supplemental Materials: Emergent topology of flat bands in a twisted bilayer $\alpha$-$T_3$ lattice}}\\[6pt]
    Gourab Paul, Srijata Lahiri, Kuntal Bhattacharyya, and Saurabh Basu\\[4pt]
    \textit{Department of Physics, Indian Institute of Technology Guwahati, Guwahati-781039, Assam, India}
\end{center}

\vspace{0.5cm}
\tableofcontents

\section{The tight-binding model of $\alpha-T_3$ lattice}
\begin{figure}[h]
    \centering
    \includegraphics[width=0.25\linewidth]{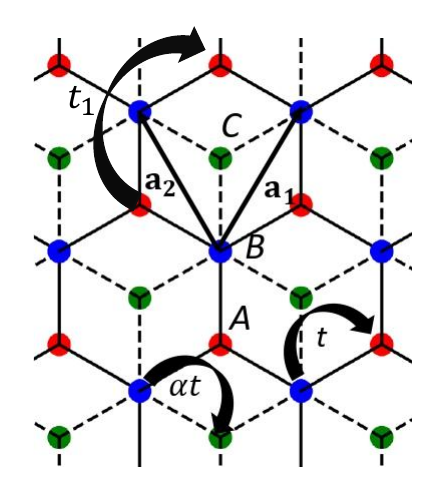}
    \caption{Schematic diagram of the $\alpha$–$T_3$ lattice. Red, blue, and green dots represent the $A$, $B$, and $C$ sublattice atoms, respectively. The primitive lattice vectors are denoted by $\mathbf{a}_1$ and $\mathbf{a}_2$. The NN hopping amplitudes between $A$–$B$ and $B$–$C$ sites are labeled as $t$ and $\alpha t$, respectively, while the NNN hopping amplitude among the $A$–$A$, $B$–$B$, and $C$–$C$ sites is represented by $t_1$.}
    \label{fig:Figure-7}
\end{figure}
The Bravais lattice vectors of the $\alpha-T_3$ lattice are given by $\mathbf{a}_1 = a \left( \frac{1}{2}, \frac{\sqrt{3}}{2} \right)$ and $\mathbf{a}_2 = a \left( -\frac{1}{2}, \frac{\sqrt{3}}{2} \right)$, where $a = \sqrt{3}d$ is the lattice constant and $d$ is the nearest-neighbor (NN) bond length. As in graphene, we set $d = 1.42$~\AA. The nearest-neighbor hopping amplitude between the $A$ and $B$ sublattices is denoted by $t$,
while that between the $B$ and $C$ sublattices is $\alpha t$, with $\alpha \in [0,1]$ interpolating between the graphene limit ($\alpha = 0$) and the dice lattice limit ($\alpha = 1$). The hopping amplitude for next-nearest-neighbor (NNN) interactions between the $A$-$A$, $B$-$B$, and $C$-$C$ sublattices is denoted by $t_1$. The schematic representation of the $\alpha-T_3$ lattice is shown in Fig.~(\ref{fig:Figure-7}). We take the $B$ site as the origin. The relative position vectors of the three nearest-neighbor $A$ sites with respect to the $B$ site are given by:
$\boldsymbol{\tau}^A_1 = (0, -d) = -\frac{1}{3}(\mathbf{a}_1 + \mathbf{a}_2)$,
$\boldsymbol{\tau}^A_2 = \left( \frac{d\sqrt{3}}{2}, \frac{d}{2} \right) = \boldsymbol{\tau}^A_1 + \mathbf{a}_1$, and
$\boldsymbol{\tau}^A_3 = \left( -\frac{d\sqrt{3}}{2}, \frac{d}{2} \right) = \boldsymbol{\tau}^A_1 + \mathbf{a}_2$.
The three nearest-neighbor vectors from the $B$ site to the $C$ sites are simply the negative of those to the $A$ sites:
$\boldsymbol{\tau}^C_i = -\boldsymbol{\tau}^A_i$ for $i = 1, 2, 3$.
The corresponding reciprocal lattice vectors are given by,
$\mathbf{b}_1 = \frac{4\pi}{\sqrt{3}a} \left( \frac{\sqrt{3}}{2}, \frac{1}{2} \right)$ and
$\mathbf{b}_2 = \frac{4\pi}{\sqrt{3}a} \left( -\frac{\sqrt{3}}{2}, \frac{1}{2} \right)$.
The $K$ and $K'$ valleys are located at $\left( \frac{4\pi}{3a}, 0 \right)$ and $\left( -\frac{4\pi}{3a}, 0 \right)$, respectively.
\par In absence of NNN hopping $t_1$, the momentum-space Hamiltonian in the sublattice basis $\left( A_{\mathbf{k}} \quad B_{\mathbf{k}} \quad C_{\mathbf{k}} \right)^{T}$ is given by~\cite{Illes2017}
\begin{equation}
    H(\mathbf{k}) =
\begin{pmatrix}
0 & f(\mathbf{k}) & 0 \\
f^*(\mathbf{k}) & 0 & \alpha f(\mathbf{k}) \\
0 & \alpha f^*(\mathbf{k}) & 0
\end{pmatrix} \label{alpha_t_3_Ham},
\end{equation}
where the off-diagonal term is defined as, $f(\mathbf{k}) = - t(1 + e^{-i \mathbf{k}.\mathbf{a}_1} + e^{-i\mathbf{k}.\mathbf{a}_2})$. Near the $K$ valley, the low-energy expansion of $f(\textbf{k})$ yields~\cite{Raoux2014, Illes2015}
\[
f(q_x, q_y) = \frac{\sqrt{3} a t}{2} (q_x - i q_y) = v_F (q_x - i q_y),
\]
where $v_F = \frac{\sqrt{3} a t}{2}$ is the Fermi velocity and $\mathbf{q}$ is measured relative to the $K$ point. Substituting this form into Eq.~\eqref{alpha_t_3_Ham}, we obtain the effective low-energy Hamiltonian
\begin{equation}
H^\prime(q_x,q_y) = v_F
\begin{pmatrix}
0 & q_x - i q_y & 0 \\
q_x + i q_y & 0 & \alpha (q_x - i q_y) \\
0 & \alpha (q_x + i q_y) & 0
\end{pmatrix}.
\end{equation}
Now including the NNN hopping amplitude $t_1$, the effective low-energy Hamiltonian near the $K$ valley can be written as
\begin{equation}
H(q_x,q_y) = v_F
\begin{pmatrix}
0 & q_x - i q_y & 0 \\
q_x + i q_y & 0 & \alpha (q_x - i q_y) \\
0 & \alpha (q_x + i q_y) & 0
\end{pmatrix} + t_2 \begin{pmatrix}
q^2 & 0 & 0 \\
0 & q^2 & 0 \\
0 & 0 & q^2
\end{pmatrix}
\end{equation}
where $t_2 = \frac{3}{4}t_1a^2 = \frac{\sqrt{3} a t_1}{2 t} v_F$ and $q^2 = q_x^2+q_y^2$.
\section{The Interlayer Hamiltonian of twisted bilayer $\alpha-T_3$ lattice}
In the vicinity of $K$, the matrix elements of the interlayer Hamiltonian are given ~\cite{Koshino2015, Ma2024}
\begin{equation}
    U_{\tilde{X},X}(\mathbf{q}, \tilde{\mathbf{q}}) = \sum_{\mathbf{G}, \tilde{\mathbf{G}}} w_{\tilde{X},X}(\mathbf{q} + \mathbf{K} + \mathbf{G}) \, e^{-i \mathbf{G} \cdot \boldsymbol{\tau}_X + i \tilde{\mathbf{G}} \cdot \boldsymbol{\tau}_{\tilde{X}}}
    \delta_{\mathbf{q} + \mathbf{K} + \mathbf{G}, \tilde{\mathbf{q}} + \tilde{\mathbf{K}} + \tilde{\mathbf{G}}}. \label{Interlayer_Ham}
\end{equation}
Here, $\mathbf{G} = m_1 \mathbf{b}_1 + m_2 \mathbf{b}_2$ and $\tilde{\mathbf{G}} = m_1 \tilde{\mathbf{b}}_1 + m_2 \tilde{\mathbf{b}}_2$, where $\mathbf{b}_1$ ($\tilde{\mathbf{b}}_1$) and $\mathbf{b}_2$ ($\tilde{\mathbf{b}}_2$) are the reciprocal lattice vectors of layer-1 (layer-2), corresponding to its primitive lattice vectors  $\mathbf{a}_1$ ($\tilde{\mathbf{a}}_1$) and $\mathbf{a}_2$ ($\tilde{\mathbf{a}}_2$).
Now, we consider the A–B (Bernal) stacking configuration~\cite{Ma2024}, under which the sublattice position vector of the atoms in both layers are given as, $\mathbf{\tau}_A = -\frac{1}{3}\left( \mathbf{a}_1 + \mathbf{a}_2\right)$, $\mathbf{\tau}_B = 0$, $\mathbf{\tau}_C = \frac{1}{3}\left( \mathbf{a}_1 + \mathbf{a}_2\right)$, $\mathbf{\tau}_{\tilde{A}} = \mathbf{\tau}_0 + s \hat{e}_z + \frac{1}{3} \left( \tilde{\mathbf{a}}_1 + \tilde{\mathbf{a}}_2\right)$, $\mathbf{\tau}_{\tilde{B}} = \mathbf{\tau}_0 + s \hat{e}_z - \frac{1}{3} \left( \tilde{\mathbf{a}}_1 + \tilde{\mathbf{a}}_2\right)$, and $\mathbf{\tau}_{\tilde{C}} = \mathbf{\tau}_0 + s \hat{e}_z$. Here, $s$ denotes the interlayer separation, and $\mathbf{\tau}_0$ is the relative in-plane translation vector of layer-2 with respect to layer-1. For simplicity, we consider $\mathbf{\tau}_0 = 0$ in our analysis. Now replacing $\mathbf{G}$ and $\tilde{\mathbf{G}}$ in Eq.~\eqref{Interlayer_Ham}, the interlayer Hamiltonian takes the form
\begin{equation}
    U_{\tilde{X},X}(\mathbf{q}, \tilde{\mathbf{q}}) = \sum_{m_1, m_2} w_{\tilde{X},X}(\mathbf{q} + \mathbf{K} + m_1 \mathbf{b}_1 + m_2 \mathbf{b}_2) \, e^{-i (m_1 \mathbf{b}_1 + m_2 \mathbf{b}_2). \boldsymbol{\tau}_X + i (m_1 \tilde{\mathbf{b}}_1 + m_2 \tilde{\mathbf{b}}_2) . \boldsymbol{\tau}_{\tilde{X}}}
    \delta_{\mathbf{q} + \mathbf{K} + m_1 \mathbf{b}_1 + m_2 \mathbf{b}_2, \tilde{\mathbf{q}} + \tilde{\mathbf{K}} + m_1 \tilde{\mathbf{b}}_1 + m_2 \tilde{\mathbf{b}}_2}. \label{Interlayer_Ham_after}
\end{equation}
The term $\delta_{\mathbf{q} + \mathbf{K} + m_1 \mathbf{b}_1 + m_2 \mathbf{b}_2, \tilde{\mathbf{q}} + \tilde{\mathbf{K}} + m_1 \tilde{\mathbf{b}}_1 + m_2 \tilde{\mathbf{b}}2}$ in Eq.~\eqref{Interlayer_Ham_after} implies that interlayer scattering between layer-1 and layer-2 is only allowed for specific combinations of $(m_1, m_2)$, for which the coupling $w_{X,\tilde{X}}(\mathbf{q} + \mathbf{K} + m_1 \mathbf{b}_1 + m_2 \mathbf{b}_2)$ is maximized.
The scattering matrices $U_{\mathbf{q}_b}$, $U_{\mathbf{q}_{tr}}$, and $U_{\mathbf{q}_{tl}}$ correspond to momentum transfers $\mathbf{q}_b$, $\mathbf{q}_{tr}$, and $\mathbf{q}_{tl}$, arising from the combinations of $(m_1, m_2) = (0,0),\ (0,1),\ \text{and}\ (-1,0)$, respectively. The corresponding matrix elements are given by
\begin{eqnarray}
    U_{\tilde{A},A} &=& w_{\tilde{A},A}(\mathbf{q} + \mathbf{K}) \delta_{\mathbf{q}-\tilde{\mathbf{q}}-\mathbf{q}_b} + w_{\tilde{A},A}(\mathbf{q} + \mathbf{K} + \mathbf{b}_2) e^{i2\pi/3} \delta_{\mathbf{q}-\tilde{\mathbf{q}}-\mathbf{q}_{tr}} + w_{\tilde{A},A}(\mathbf{q}+\mathbf{K} - \mathbf{b}_1) e^{-i2\pi/3} \delta_{\mathbf{q}-\tilde{\mathbf{q}}-\mathbf{q}_{tl}}\nonumber\\ \nonumber\\
    &=& w_1 \delta_{\mathbf{q}-\tilde{\mathbf{q}}-\mathbf{q}_b} + w_1 e^{i2\pi/3} \delta_{\mathbf{q}-\tilde{\mathbf{q}}-\mathbf{q}_{tr}} + w_1 e^{-i2\pi/3} \delta_{\mathbf{q}-\tilde{\mathbf{q}}-\mathbf{q}_{tl}}\\\nonumber\\
    U_{\tilde{A},B} &=& w_{\tilde{A},B}(\mathbf{q} + \mathbf{K}) \delta_{\mathbf{q}-\tilde{\mathbf{q}}-\mathbf{q}_b} + w_{\tilde{A},B}(\mathbf{q} + \mathbf{K} + \mathbf{b}_2) e^{-i2\pi/3} \delta_{\mathbf{q}-\tilde{\mathbf{q}}-\mathbf{q}_{tr}} + w_{\tilde{A},B}(\mathbf{q}+\mathbf{K} - \mathbf{b}_1) e^{i2\pi/3} \delta_{\mathbf{q}-\tilde{\mathbf{q}}-\mathbf{q}_{tl}} \nonumber\\ \nonumber\\
    &=& w_2 \delta_{\mathbf{q}-\tilde{\mathbf{q}}-\mathbf{q}_b} + w_2 e^{-i2\pi/3} \delta_{\mathbf{q}-\tilde{\mathbf{q}}-\mathbf{q}_{tr}} + w_2 e^{i2\pi/3} \delta_{\mathbf{q}-\tilde{\mathbf{q}}-\mathbf{q}_{tl}}\\\nonumber\\
    U_{\tilde{A},C} &=& w_{\tilde{A},C}(\mathbf{q} + \mathbf{K}) \delta_{\mathbf{q}-\tilde{\mathbf{q}}-\mathbf{q}_b} + w_{\tilde{A},C}(\mathbf{q} + \mathbf{K} + \mathbf{b}_2) \delta_{\mathbf{q}-\tilde{\mathbf{q}}-\mathbf{q}_{tr}} + w_{\tilde{A},C}(\mathbf{q}+\mathbf{K} - \mathbf{b}_1) \delta_{\mathbf{q}-\tilde{\mathbf{q}}-\mathbf{q}_{tl}} \nonumber\\ \nonumber\\
    &=& w_3 \delta_{\mathbf{q}-\tilde{\mathbf{q}}-\mathbf{q}_b} + w_3 \delta_{\mathbf{q}-\tilde{\mathbf{q}}-\mathbf{q}_{tr}} + w_3 \delta_{\mathbf{q}-\tilde{\mathbf{q}}-\mathbf{q}_{tl}}\\ \nonumber\\
    U_{\tilde{B},A} &=& w_{\tilde{B},A}(\mathbf{q} + \mathbf{K}) \delta_{\mathbf{q}-\tilde{\mathbf{q}}-\mathbf{q}_b} + w_{\tilde{B},A}(\mathbf{q} + \mathbf{K} + \mathbf{b}_2) \delta_{\mathbf{q}-\tilde{\mathbf{q}}-\mathbf{q}_{tr}} + w_{\tilde{B},A}(\mathbf{q}+\mathbf{K} - \mathbf{b}_1) \delta_{\mathbf{q}-\tilde{\mathbf{q}}-\mathbf{q}_{tl}} \nonumber\\ \nonumber\\
    &=& w_2 \delta_{\mathbf{q}-\tilde{\mathbf{q}}-\mathbf{q}_b} + w_2 \delta_{\mathbf{q}-\tilde{\mathbf{q}}-\mathbf{q}_{tr}} + w_2 \delta_{\mathbf{q}-\tilde{\mathbf{q}}-\mathbf{q}_{tl}}\\\nonumber\\
    U_{\tilde{B},B} &=& w_{\tilde{B},B}(\mathbf{q} + \mathbf{K}) \delta_{\mathbf{q}-\tilde{\mathbf{q}}-\mathbf{q}_b} + w_{\tilde{B},B}(\mathbf{q} + \mathbf{K} + \mathbf{b}_2) e^{i2\pi/3}\delta_{\mathbf{q}-\tilde{\mathbf{q}}-\mathbf{q}_{tr}} + w_{\tilde{B},B}(\mathbf{q}+\mathbf{K} - \mathbf{b}_1) e^{-i2\pi/3}\delta_{\mathbf{q}-\tilde{\mathbf{q}}-\mathbf{q}_{tl}} \nonumber\\ \nonumber\\
    &=& w_1 \delta_{\mathbf{q}-\tilde{\mathbf{q}}-\mathbf{q}_b} + w_1 e^{i2\pi/3}\delta_{\mathbf{q}-\tilde{\mathbf{q}}-\mathbf{q}_{tr}} + w_1 e^{-i2\pi/3}\delta_{\mathbf{q}-\tilde{\mathbf{q}}-\mathbf{q}_{tl}}\\\nonumber\\
    U_{\tilde{B},C} &=& w_{\tilde{B},C}(\mathbf{q} + \mathbf{K}) \delta_{\mathbf{q}-\tilde{\mathbf{q}}-\mathbf{q}_b} + w_{\tilde{B},C}(\mathbf{q} + \mathbf{K} + \mathbf{b}_2) e^{-i2\pi/3}\delta_{\mathbf{q}-\tilde{\mathbf{q}}-\mathbf{q}_{tr}} + w_{\tilde{B},C}(\mathbf{q}+\mathbf{K} - \mathbf{b}_1) e^{i2\pi/3}\delta_{\mathbf{q}-\tilde{\mathbf{q}}-\mathbf{q}_{tl}} \nonumber\\ \nonumber\\
    &=& w_2 \delta_{\mathbf{q}-\tilde{\mathbf{q}}-\mathbf{q}_b} + w_2 e^{-i2\pi/3}\delta_{\mathbf{q}-\tilde{\mathbf{q}}-\mathbf{q}_{tr}} + w_2 e^{i2\pi/3}\delta_{\mathbf{q}-\tilde{\mathbf{q}}-\mathbf{q}_{tl}}\\\nonumber\\
    U_{\tilde{C},A} &=& w_{\tilde{C},A}(\mathbf{q} + \mathbf{K}) \delta_{\mathbf{q}-\tilde{\mathbf{q}}-\mathbf{q}_b} + w_{\tilde{C},A}(\mathbf{q} + \mathbf{K} + \mathbf{b}_2) e^{-i2\pi/3}\delta_{\mathbf{q}-\tilde{\mathbf{q}}-\mathbf{q}_{tr}} + w_{\tilde{C},A}(\mathbf{q}+\mathbf{K} - \mathbf{b}_1) e^{i2\pi/3}\delta_{\mathbf{q}-\tilde{\mathbf{q}}-\mathbf{q}_{tl}} \nonumber\\ \nonumber\\
    &=& w_3 \delta_{\mathbf{q}-\tilde{\mathbf{q}}-\mathbf{q}_b} + w_3 e^{-i2\pi/3}\delta_{\mathbf{q}-\tilde{\mathbf{q}}-\mathbf{q}_{tr}} + w_3 e^{i2\pi/3}\delta_{\mathbf{q}-\tilde{\mathbf{q}}-\mathbf{q}_{tl}}\\\nonumber\\
    U_{\tilde{C},B} &=& w_{\tilde{C},B}(\mathbf{q} + \mathbf{K}) \delta_{\mathbf{q}-\tilde{\mathbf{q}}-\mathbf{q}_b} + w_{\tilde{C},B}(\mathbf{q} + \mathbf{K} + \mathbf{b}_2) \delta_{\mathbf{q}-\tilde{\mathbf{q}}-\mathbf{q}_{tr}} + w_{\tilde{C},B}(\mathbf{q}+\mathbf{K} - \mathbf{b}_1) \delta_{\mathbf{q}-\tilde{\mathbf{q}}-\mathbf{q}_{tl}}\nonumber\\ \nonumber\\
    &=& w_2 \delta_{\mathbf{q}-\tilde{\mathbf{q}}-\mathbf{q}_b} + w_2  \delta_{\mathbf{q}-\tilde{\mathbf{q}}-\mathbf{q}_{tr}} + w_2 \delta_{\mathbf{q}-\tilde{\mathbf{q}}-\mathbf{q}_{tl}}\\\nonumber\\
    U_{\tilde{C},C} &=& w_{\tilde{C},C}(\mathbf{q} + \mathbf{K}) \delta_{\mathbf{q}-\tilde{\mathbf{q}}-\mathbf{q}_b} + w_{\tilde{C},C}(\mathbf{q} + \mathbf{K} + \mathbf{b}_2) e^{i2\pi/3} \delta_{\mathbf{q}-\tilde{\mathbf{q}}-\mathbf{q}_{tr}} + w_{\tilde{C},C}(\mathbf{q}+\mathbf{K} - \mathbf{b}_1) e^{-i2\pi/3} \delta_{\mathbf{q}-\tilde{\mathbf{q}}-\mathbf{q}_{tl}}\nonumber\\ \nonumber\\
    &=& w_3 \delta_{\mathbf{q}-\tilde{\mathbf{q}}-\mathbf{q}_b} + w_3 e^{i2\pi/3} \delta_{\mathbf{q}-\tilde{\mathbf{q}}-\mathbf{q}_{tr}} + w_3 e^{-i2\pi/3} \delta_{\mathbf{q}-\tilde{\mathbf{q}}-\mathbf{q}_{tl}}
\end{eqnarray}
Now the interlayer Hamiltonian is given by
\begin{eqnarray}
U(\mathbf{q},\tilde{\mathbf{q}}) 
&=& 
\begin{pmatrix}
U_{\tilde{A},A} & U_{\tilde{B},A} & U_{\tilde{C},A}\\
U_{\tilde{A},B} & U_{\tilde{B},B} & U_{\tilde{C},B}\\
U_{\tilde{A},C}& U_{\tilde{B},C} & U_{\tilde{C},C} \nonumber
\end{pmatrix}\\ \nonumber \\
&=&
\begin{pmatrix}
w_1 & w_2 & w_3\\
w_2 & w_1 & w_2 \\
w_3 & w_2 & w_1
\end{pmatrix}\delta_{\mathbf{q}-\tilde{\mathbf{q}}-\mathbf{q}_b} + 
\begin{pmatrix}
w_1 e^{i\phi}& w_2 & w_3e^{-i\phi}\\
w_2e^{-i\phi} & w_1 e^{i\phi} & w_2 \\
w_3 & w_2 e^{-i\phi}& w_1e^{i\phi}
\end{pmatrix}  \delta_{\mathbf{q}-\tilde{\mathbf{q}}-\mathbf{q}_{tr}}+
\begin{pmatrix}
w_1 e^{-i\phi}& w_2 & w_3e^{i\phi}\\
w_2e^{i\phi} & w_1 e^{-i\phi} & w_2 \\
w_3 & w_2 e^{i\phi}& w_1e^{-i\phi}
\end{pmatrix} \delta_{\mathbf{q}-\tilde{\mathbf{q}}-\mathbf{q}_{tl}}\nonumber\\ \nonumber\\
&=& U_{\mathbf{q}_b} \delta_{\mathbf{q}-\tilde{\mathbf{q}}-\mathbf{q}_b} + U_{\mathbf{q}_{tr}} \delta_{\mathbf{q}-\tilde{\mathbf{q}}-\mathbf{q}_{tr}} + U_{\mathbf{q}_{tl}} \delta_{\mathbf{q}-\tilde{\mathbf{q}}-\mathbf{q}_{tl}} \label{Interlayer_Coupling}
\end{eqnarray}
The matrices $U_{\mathbf{q}_b}$, $U_{\mathbf{q}_{tr}}$, $U_{\mathbf{q}_{tl}}$, which describe the interlayer coupling associated with each momentum transfer, take the following forms: 
$U_{\mathbf{q}_b} =  \begin{pmatrix}
w_1 & w_2 & w_3\\
w_2 & w_1 & w_2 \\
w_3 & w_2 & w_1
\end{pmatrix},$ 
$U_{\mathbf{q}_{tr}} =  \begin{pmatrix}
w_1 e^{i\phi}& w_2 & w_3e^{-i\phi}\\
w_2e^{-i\phi} & w_1 e^{i\phi} & w_2 \\
w_3 & w_2 e^{-i\phi}& w_1e^{i\phi}
\end{pmatrix},$
$U_{\mathbf{q}_{tl}} =  \begin{pmatrix}
w_1 e^{-i\phi}& w_2 & w_3e^{i\phi}\\
w_2e^{i\phi} & w_1 e^{-i\phi} & w_2 \\
w_3 & w_2 e^{i\phi}& w_1e^{-i\phi}
\end{pmatrix}$ (with $\phi = 2\pi/3$). Eq.~\eqref{Interlayer_Coupling} yields the form of the interlayer Hamiltonian given in Eq.~(4) of our manuscript.
\section{Evolution of the hybrid Wannier charge center}
To begin with, the Wannier functions (WFs) represent an orthogonal basis, maximally localized about a real space co-ordinate (say $\mathbf{R}$), with respect to all relevant spatial dimensions \cite{WF},
\begin{eqnarray}
{|W_n(\bf R)\rangle}=\frac{V}{(2\pi)^D}\int_{BZ} d^Dk~e^{-i\bf k.\bf R}|\psi_{n\bf k}\rangle.
\label{WF}
\end{eqnarray}
Here, $|\psi_{n, \mathbf{k}}\rangle$ represents the Bloch wavefunctions, with $D$, $V$ and $n$ representing the dimensionality, the real space volume and band index respectively.
Consequently, the Wannier charge centers (WCCs) which represent the center of charge in a unit cell, can be mathematically defined as the expectation value of the position operators with respect to the localized Wannier functions.
However, while exponentially localized Wannier functions can be successfully obtained for topologically trivial systems, the existence of topological non-triviality poses a constraint in their construction in two and three dimensions.
Nevertheless, such a constraint is absent in one-dimensional systems, thereby allowing us to construct hybrid (effectively one dimensional) WFs where they are localized along one spatial direction, while being Bloch-like along the rest.
Mathematically, a hybrid WF can be defined as \cite{HWCC1},
\begin{eqnarray}
{|W_{n}^H(R_x, k_y,k_z)\rangle}=\frac{1}{(2\pi)}\int_{-\pi}^\pi dk_x~e^{-ik_xR_x}|\psi_{n\bf k}\rangle,
\label{WF}
\end{eqnarray}
where the direction of localization is $x$, with the WF being delocalized along $y$ and $z$.
The hybrid WCC (HWCC), which now represents the center of charge along the direction in which the WF is localized, is given as \cite{Taherinejad2014},
\begin{eqnarray}
{\langle x_{n}(k_y,k_z)\rangle}=\langle W_{n}^H(R_x, k_y,k_z)|\hat{\bf X}|W_{n}^H(R_x, k_y,k_z)\rangle.
\label{WF}
\end{eqnarray}
According to the modern theory of polarization, the HWCC can be interpreted in terms of the Berry phase, $\phi_n(k_y,k_z)$ as
\begin{equation}
{\langle x_{n}(k_y,k_z)\rangle}=\frac{\phi_n(k_y,k_z)}{2\pi}=\frac{1}{2\pi}\int_0^{2\pi}A_n(k_x,k_y,k_z)dk_x,
\label{WF}
\end{equation}
where $A_n(k_x,k_y,k_z)=-i\langle u_{n\bf k}|\nabla_{\bf k}|u_{n\bf k}\rangle$ denotes the Berry connection and $|u_{n\bf k}\rangle$ refers to the cell-periodic part of the Bloch wavefunction $|\psi_{n\bf k}\rangle$. 
The evolution of the HWCC as a function of the remaining periodic parameters ($k_y$ and $k_z$ for the aforementioned example), determines the topology of the system where a trivial (non-trivial) winding of the WCC corresponds to a $C=0$ ($C\neq0$) phase, as shown in Fig. 4 in the manuscript.
For the hexagonal lattice system studied in our work, a general vector in the momentum space can be written as $\mathbf{q}=q_{\hat{{\mathbf{b}}}_1^m}\hat{\mathbf{b}}_1^m + q_{\hat{\mathbf{b}}_2^m}\hat{\mathbf{b}}_2^m$, where $\hat{\mathbf b}_1^m$ and $\hat{\mathbf b}_2^m$ correspond to moiré reciprocal lattice vector directions mentioned in the manuscript.
In the evolution of HWCC shown in Fig. 4, the WCC is localized along the direction $\hat{\mathbf{a}}_1^m$, whereas $q_{\hat{\mathbf{b}}_2^m}$ is retained as a periodic parameter.
The evolution is clearly trivial as shown in Fig. 4 (a), whereas it corresponds to a phase with $|C|=1$ in Fig. 4 (b) as shown in the manuscript.\\
It is to be noted that the evolution of the hybrid Wannier charge center corresponds to the edge manifestations on a nanoribbon configuration of a system.  The evolution of the hybrid WCC with respect to a periodic parameter of the Hamiltonian, emulates the edge of a nanoribbon configuration (periodic along one direction and aperiodic along the other).
\begin{figure*}[h]
    \centering
    \includegraphics[width=0.9\textwidth]{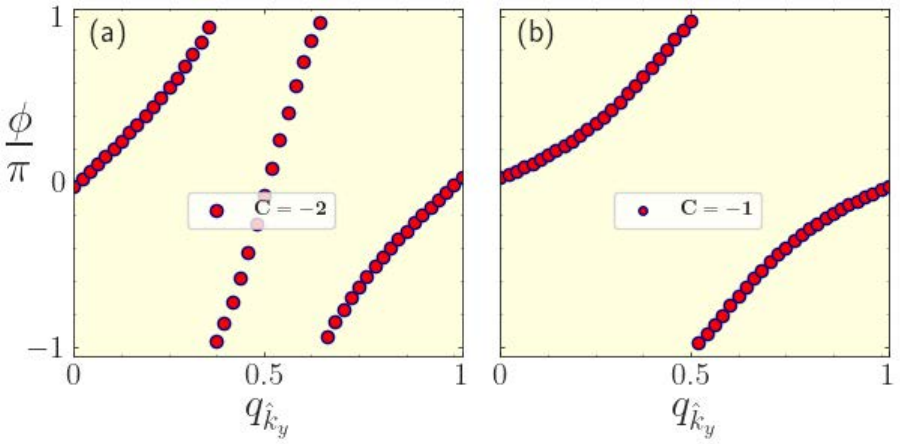}
    \caption{The evolution of the WCC corresponding to the phases (I) and (II) AS shown in Fig. \ref{fig:Figure-WCC} in the main text. The corresponding Chern number has been mentioned in the inset.}
    \label{fig:Figure-WCC}
\end{figure*}
This has been elaborately explained in Ref.~\cite{HWCC1}. For a square/rectangular BZ, the Chern number which represents the bulk topological invariant of a TRS broken topological insulator, can be written in terms of the hybrid WCC as :
\begin{equation}
C = \frac{1}{a_x}\Big{(}\sum_n \bar{x}_n(k_y=A)-\bar{x}_n(k_y=B)\Big{)}
\end{equation}
where the system has been periodized in the $y$-direction (with $A$ and $B$ representing the limits of the hybrid 1D Brillouin zone) such that $k_y$ is retained as a periodic parameter and $\bar x_n$ represents the center of charge in a unit cell along the $x$-direction. Furthermore, $n$ represents the number of occupied bands. It is to be noted that the translational symmetry of the underlying lattice together with the non-trivial Chern number ensures that after an evolution of $k_y$ over a complete period, the WCC travels by an integer number of the relevant lattice translation vector, which is in turn represented by the winding of the WCC as shown in Fig. \ref{fig:Figure-WCC} for the different topological phases of our system. {Furthermore, charge conservation ensures that the edge bandstructure has to have a state crossing the band gap if the hybrid WCC winds during the evolution along a full circle of the periodic system parameter.} This occurs to ensure that the accumulation of charge caused by the pumping of the WCC during a periodic evolution of $k_y$ (here) at one edge of the system, can vanish through the edge state into the conduction band and thereby maintain charge neutrality. Thus, we plot the evolution of the WCC as a function of momenta in one of the rectangular directions (say $k_y$ here) for the different topological phases (region (I) and (II))as shown in the phase plot (Fig. 6) in the main manuscript.\\
Furthermore, the bulk topological invariant, that is, the Chern number, is directly related to the anomalous Hall conductivity as
$C = \frac{h}{e^2}\sigma_{xy}$~\cite{Zhang2019A}. The Chern number of the non-degenerate sub-band at the boundary of the middle band, therefore, determines its contribution to the Hall conductivity of the system.\\
\begin{figure}[h]
    \centering
    \includegraphics[width=1.0\linewidth]{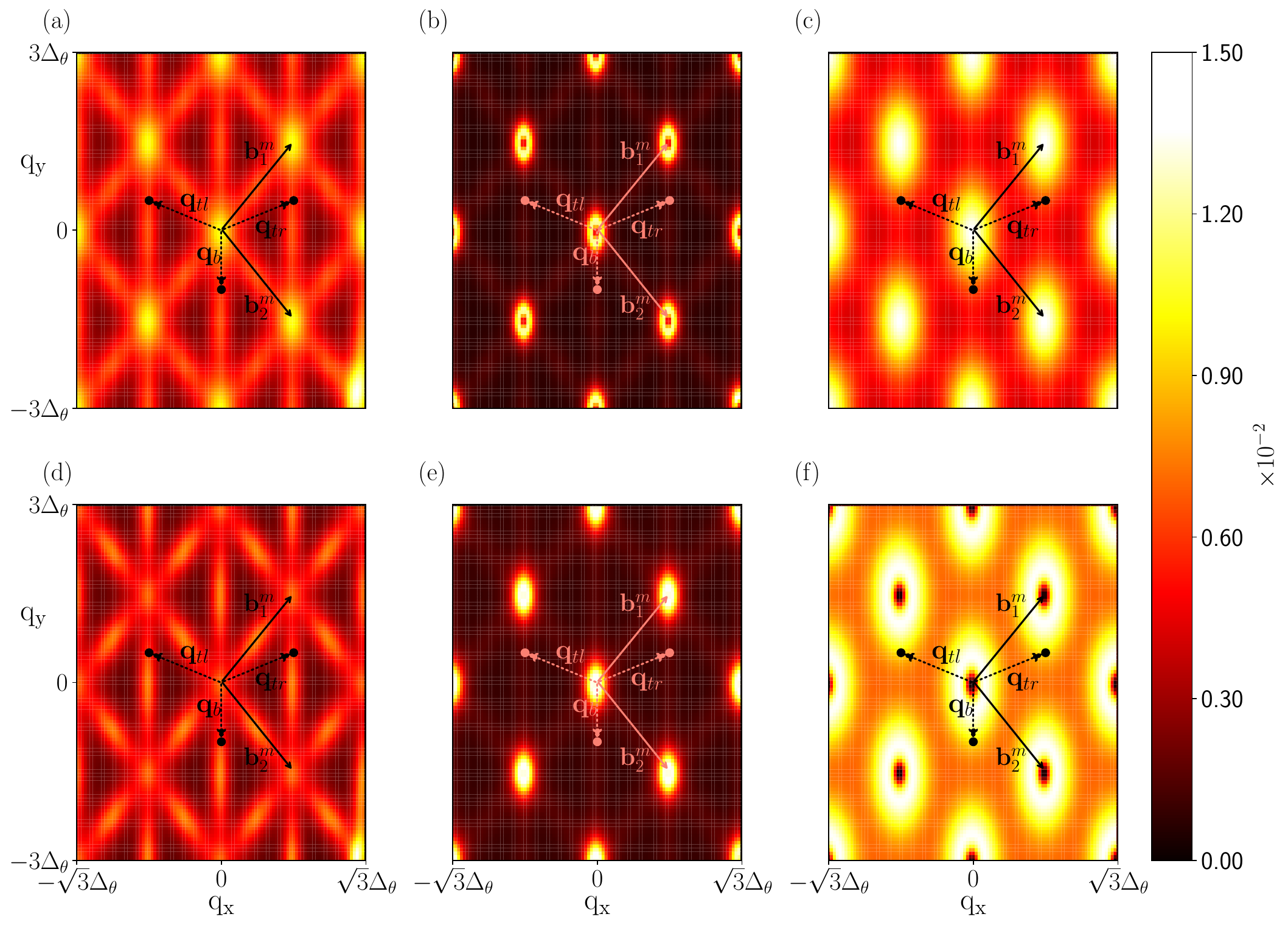}
    \caption{The Berry curvature of the NFB in twisted bilayer $\alpha$–$T_3$ lattice is plotted as a function of $q_x$ and $q_y$.Panels (a)–(c) show the Berry-curvature hotspots of the NFB in the dice limit (keeping $\alpha$ fixed at 1) for $\theta = 0.32^\circ$, $0.50^\circ$, and $0.80^\circ$, with $t_2 = 0.001~v_F$. Panels (d)–(f) display the Berry curvature at the magic angle ($\theta = 1.08^\circ$) for $\alpha = 0.30$, $0.50$, and $0.70$, respectively, with $t_2 =0.001~v_F$. The other parameters, $w_1 = 0$, $w_2 = 110.7$ meV, $w_3 = 0$, $v_F = 6326.1$ meV$\cdot$\AA, and $M = 17$ meV, are kept same for all the panels (a)-(f).}
    \label{fig:Figure-11}
\end{figure}
\section{Numerical analysis of Berry curvature and Chern number}
Along with the evolution of the band structure and HWCC of the NFB in the $\alpha-\theta$ plane, we also calculate the Berry curvature and Chern number of the NFB using Fukui’s method~\cite{Fukui2005}. In order to carry out the calculation, we discretize the MBZ into an $N_x \times N_y$ grid, with each site corresponding to a momentum point $\mathbf{q} = (q_x, q_y)$. Now for each momentum point, we diagonalize the Hamiltonian $H(\mathbf{q})$ and extract the eigenvector $\psi(\mathbf{q})$ corresponding to the NFB. Now we evaluate the following complex quantities
\begin{eqnarray}
    \zeta_{\hat{x}} (\mathbf{q}) = \langle \psi(\mathbf{q} + \delta q_x \hat{q}_x) | \psi (\mathbf{q}) \rangle
\quad \text{and} \quad
    \zeta_{\hat{y}} (\mathbf{q}) = \langle \psi(\mathbf{q} + \delta q_y \hat{q}_y) | \psi (\mathbf{q}) \rangle,
\end{eqnarray}
where $\hat{q}_x$ and $\hat{q}_y$ denote the unit vectors along the direction $q_x$ and $q_y$ in momentum space. Here, $\delta q_x$ and $\delta q_y$ are the constant separations between neighboring grid points along the direction $q_x$ and $q_y$, respectively. The Berry curvature at $\mathbf{q}$ is defined as
\begin{eqnarray}
\mathbf{B}(\mathbf{q}) = \frac{\arg W(\mathbf{q})}{\delta q_x  \delta q_y}, \label{Berry_curvature}
\end{eqnarray}
where $W(\mathbf{q})$ is the gauge-invariant quantity given by
\begin{eqnarray}
    W(\mathbf{q}) = \zeta_{\hat{x}}(\mathbf{q}) \zeta_{\hat{y}}(\mathbf{q} + \delta q_x \hat{q}_x) \zeta^*_{\hat{x}}(\mathbf{q} + \delta q_y \hat{q}_y) \zeta_{\hat{y}}^*(\mathbf{q})
\end{eqnarray}
To avoid any ambiguity in the evaluation of the Berry curvature, it is important to guarantee that $|\text{arg}~W(\mathbf{q})|< \pi$ for any $\mathbf{q}$. As long as the Berry curvature remains non-singular, this condition can always be satisfied by reducing the plaquette size. The numerical error can be reduced by increasing the number of grid points $N_x$ and $N_y$. Then the Chern number is given by 
\begin{eqnarray}
    C = \frac{1}{2\pi} \sum_{\mathbf{q}} \mathbf{B(q)} \delta q_x \delta q_y \label{Chern_number}
\end{eqnarray}
In our analysis, we take $N_x = N_y = 20$, which is sufficient to obtain accurate results for the Berry curvature and the Chern number. Now using Eq.~\eqref{Berry_curvature}, we show the Berry curvature hotspots of the NFB in Fig.~(\ref{fig:Figure-11}). We plot the Berry curvature of the NFB in two limiting cases. In the first case, corresponding to the dice limit ($\alpha = 1$), the results for $\theta = 0.32^\circ$, $0.50^\circ$, and $0.80^\circ$ are shown in Fig.~[\ref{fig:Figure-11}(a)–\ref{fig:Figure-11}(c)]. In second case, with the system fixed at the magic angle ($\theta = 1.08^\circ$), the results for increasing values of $\alpha$ ($0.3
0$, $0.50$, and $0.70$) are presented in Fig.~[\ref{fig:Figure-11}(d)–\ref{fig:Figure-11}(f)], respectively. The Chern number associated with the NFB is evaluated using Eq.~\eqref{Chern_number}. The detailed systematic analysis of the Chern number across the $\alpha-\theta$ plane, highlighting distinct topological phase transitions through gap-closing and gap-opening processes, is presented in the main text.
\begin{figure}[h]
    \centering
    \includegraphics[width=0.6\linewidth]{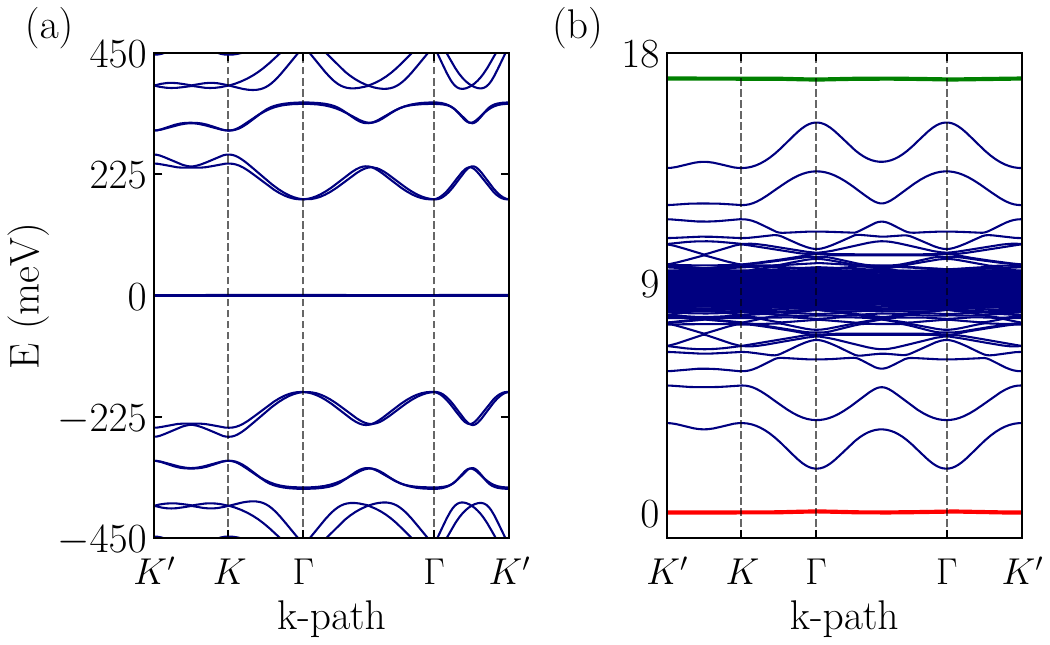}
    \caption{Band structure of the \textit{twisted bilayer} $\alpha-T_3$ lattice in the dice ($\alpha=1$) limit, where panel (a) demonstrates its variation evaluated at the magic angle $\theta = 1.08^\circ$ in the chiral limit ($w_1 = 0$, $w_2 = 110.7$~meV, $w_3 = 0$), with Fermi velocity $v_F = 6326.1$~meV$\cdot$\AA~and $t_2 = 0.001~v_F$. The panel (b) is obtained with an alignment of two h-BN substrate layers at $A$ and $B$ sublattice sites, where one layer is placed above the top $\alpha-T_3$ layer and the other beneath the bottom $\alpha-T_3$ layer. This special construction introduces a staggered sublattice potential, resulting in a broadening of the middle band and the emergence of NFBs, marked by solid green and red lines.}
    \label{fig:Figure-S4}
\end{figure}
\section{Effect of $\text{h-BN}$ alignment with A and B atoms of the twisted bilayer $\alpha-T_3$ lattice}
\begin{figure*}
    \centering
    \includegraphics[width=1.0\textwidth]{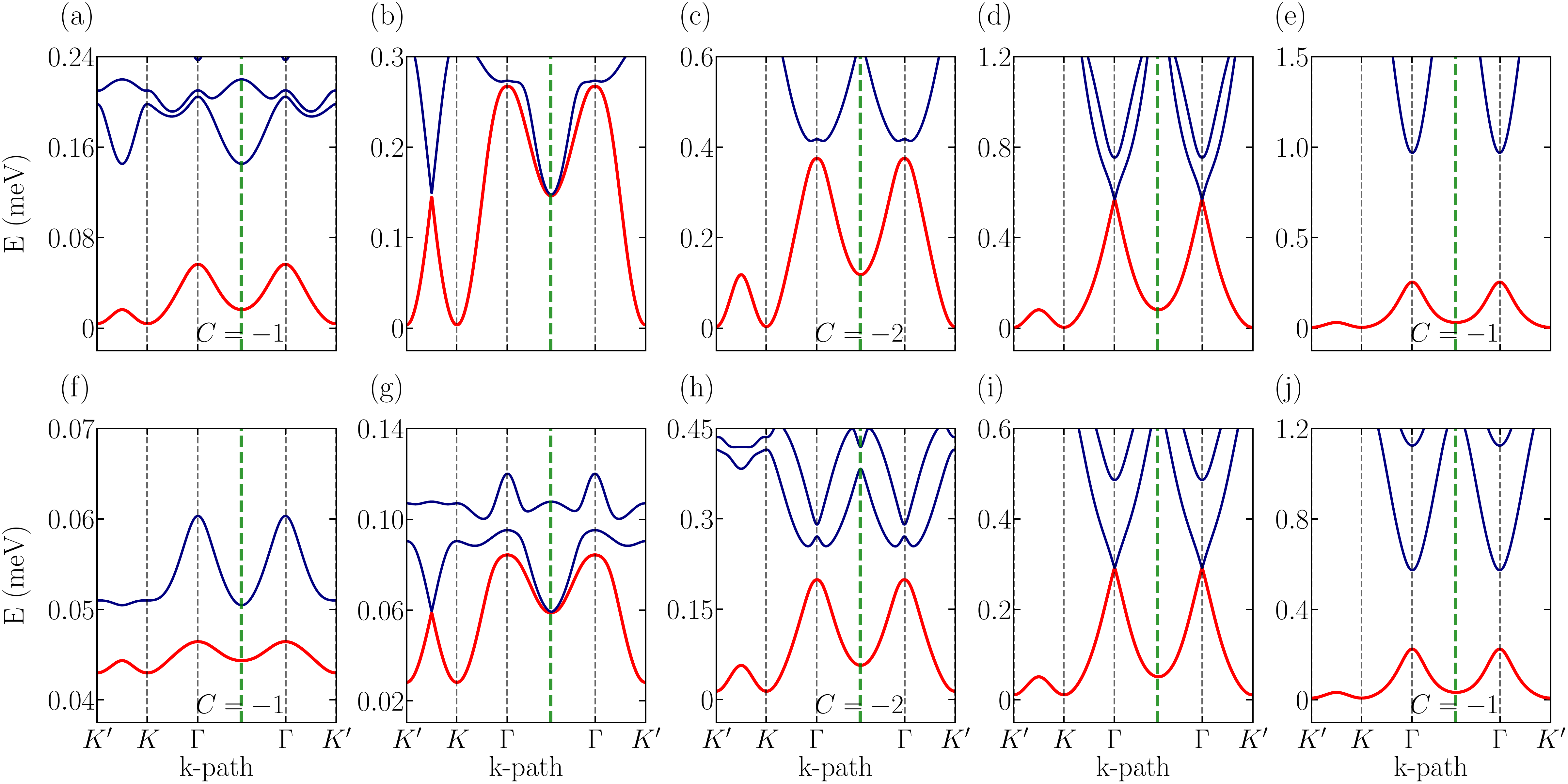}
    \caption{The band structures of the NFB in the dice limit are shown in panels (a)–(e) for increasing twist angles $\theta = 0.32^{\circ}, 0.40^{\circ}, 0.50^{\circ}, 0.62^{\circ},$ and $0.80^{\circ}$, with $t_2 = 0.001~v_F$. Panels (f)–(j) display the evolution of the NFB at fixed magic angle $\theta = 1.08^{\circ}$ for increasing values of $\alpha = 0.28, 0.357, 0.50, 0.567,$ and $0.70$, respectively, with $t_2 = 0.001~v_F$. All other parameters are kept identical to those in Fig.~\ref{fig:Figure-3}(b). The green dotted line, drawn perpendicular to the k-path between two $\Gamma$ points, marks the $M$ point along the high-symmetry path in the MBZ.}
    \label{fig:Figure-5_AB}
\end{figure*}
In this section, we discuss the effect of h-BN alignment with the $A$ and $B$ atoms in both layers of the \textit{twisted bilayer} $\alpha-T_3$ lattice. For this purpose, we adopt the similar treatment as in the case of h-BN alignment with the $A$ and $C$ atoms of the \textit{twisted bilayer} $\alpha-T_3$ lattice, as outlined in our manuscript. The key distinction of the $A$-$B$ alignment, compared to the $A$-$C$ case, is the inclusion of a different `mass' term in the intralayer part of the tight-binding Hamiltonian. Here, the `mass'
term is added such that the $A$ ($B$) sublattice is given an onsite potential of $M$ ($-M$).
Now under the combined conditions of the chiral limit ($w_1 = w_3 = 0$)~\cite{Tarnopolsky2019,Zhou2024}, the magic angle ($\theta = 1.08^\circ$), and the dice limit ($\alpha = 1$), we plot the band structure of the system, as shown in Fig.~\ref{fig:Figure-S4}. It is observed that the system hosts two isolated, non-degenerate NFBs at the edges of the broadened spectrum of the middle band. However, the symmetry of these bands about zero energy is lost, with the NFBs now located at $E = 0$ and $M$, indicated by the solid red and green lines as shown in Fig.~\ref{fig:Figure-S4}(b). 
Now in the remainder of this section, we systematically analyze the topological properties of the NFB located at $E = 0$ (which was previously at $E = -M$ for the $A$–$C$ alignment) as a function of $\alpha$ and $\theta$, characterized by the single-band Chern number ($C$) computed using Fukui’s method. 
\par Similar to the study done in the manuscript, we analyze the band structure of the NFB and its topological evolution in two limiting cases. First, in the dice limit (fixing $\alpha=1$), for a substantial range of twist angle $\theta$. Second, in the magic-angle limit (keeping $\theta$ fix at $1.08^\circ$), while varying $\alpha$ continuously over the entire range between $0$ and $1$.
It is observed (see Fig.~\ref{fig:Figure-5_AB}) that the NFB exhibits similar kind of band-gap closing and reopening scenarios as in the $A$–$C$ alignment case discussed in our manuscript. The evolution of the NFB in the dice limit with increasing $\theta$  values is shown in Fig.~\ref{fig:Figure-5_AB}(a)-\ref{fig:Figure-5_AB}(e), while the magic angle limit with increasing $\alpha$ values is shown in Fig.~\ref{fig:Figure-5_AB}(f)-\ref{fig:Figure-5_AB}(j). 
The values of the Chern number $C$, obtained using Fukui’s method, are explicitly indicated in the corresponding panels, illustrating how the band topology of the NFB evolves with $\alpha$ and $\theta$ in the two limiting cases.
\section{Impact of chiral symmetry breaking on the band structure and topological evolution of the NFB}
\begin{figure*}
    \centering
    \includegraphics[width=1.0\textwidth]{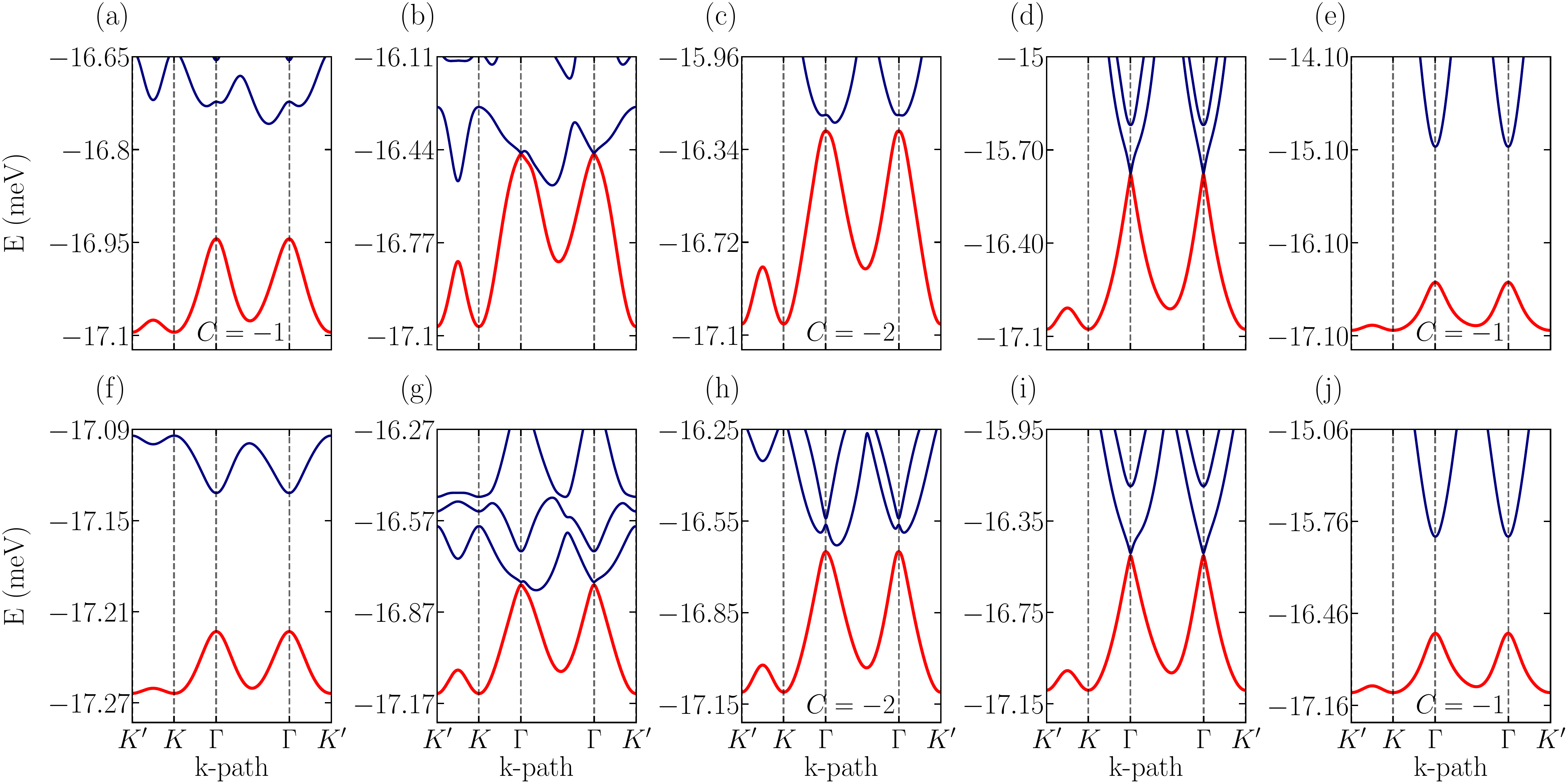}
    \caption{Panels (a)–(e) illustrate the band structures of the NFB in the dice limit for progressively increasing twist angles, $\theta = 0.32^{\circ}, 0.42^{\circ}, 0.50^{\circ}, 0.62^{\circ},$ and $0.80^{\circ}$, using parameters $t_2 = 0.001~v_F$, $w_1 = w_3 = 0.003~w_2$, and $w_2 = 110.7$~meV. In contrast, panels (f)–(j) depict the evolution of the NFB at a fixed magic angle $\theta = 1.08^{\circ}$ as $\alpha$ increases through values $0.30, 0.448, 0.50, 0.556,$ and $0.70$, respectively, with $t_2 = 0.001~v_F$, $w_1 = w_3 = 0.003~w_2$, and $w_2 = 110.7$~meV.}
    \label{fig:Figure-S6}
\end{figure*}
In this section, we examine how finite values of $w_1$ and $w_3$ influence the band structure of the NFB and their impact on its topological evolution. To examine this, we consider h-BN alignment with the $A$ and $C$ atoms in both layers of the \textit{twisted bilayer} $\alpha-T_3$ lattice. We begin by examining the band structure of the NFB in the dice limit ($\alpha = 1$) over a broad range of twist angles $\theta$, considering $t_2 = 0.001~ v_F$, $w_1 = w_3 = 0.003~w_2$ and $w_2 = 110.7$~meV. The corresponding band structures for increasing $\theta$ values ($0.32^\circ$, $0.42^\circ$, $0.50^\circ$, $0.62^\circ$, and $0.80^\circ$) are shown in Figs.~[\ref{fig:Figure-S6}(a)-\ref{fig:Figure-S6}(e)]. 
At $\theta = 0.32^\circ$, the NFB lies in the topological phase characterized by the Chern number, $C = -1$ [Fig.~\ref{fig:Figure-S6}(a)]. 
As we increase the twist angle $\theta$, this NFB undergoes a first order topological phase transition from $C=-1$ at $\theta=0.32^\circ$ to $C=-2$ at $\theta=0.50^\circ$ [Fig.~\ref{fig:Figure-S6}(c)], via a gap closing transition at $\theta=0.42^\circ$ [Fig.~\ref{fig:Figure-S6}(b)], In absence of $w_1$ and $w_3$, this gap closing point was previously at $\theta=0.40^\circ$. On further increasing $\theta$, the system undergoes another gap closing event occurring at $\theta = 0.62^\circ$ [Fig.~\ref{fig:Figure-S6}(d)], which leads to another first order topological phase transition of the NFB from $C=-2$ back to $C=-1$ [Fig.~\ref{fig:Figure-S6}(e)] at the $\Gamma$ point of the MBZ. At $\theta = 1.08^\circ$ (in the dice limit), the NFB shifts from $E = -M$ but the band resides in the $C = -1$ topological phase. Now we investigate the electronic structure of the NFB at the magic angle, fixing $\theta = 1.08^\circ$, and setting the parameters as $t_2 = 0.001~v_F$, $w_1 = w_3 = 0.003~w_2$, and $w_2 = 110.7$~meV, while exploring the entire range of $\alpha$ between $0$ and $1$. The corresponding band structures for increasing values of $\alpha$ (0.30, 0.448, 0.50, 0.556, and 0.70) are illustrated in Fig.~[\ref{fig:Figure-S6}(f)–\ref{fig:Figure-S6}(j)]. It is observed that for larger values of $\alpha$, the topological phases remain robust, whereas for smaller $\alpha$ values ($\alpha < 0.47$), the Chern number of the NFB becomes ill-defined. The band structures for $\alpha = 0.30$ and $\alpha = 0.448$ are presented in Fig.~\ref{fig:Figure-S6}(f) and Fig.~\ref{fig:Figure-S6}(g), respectively. Although a gap-closing transition occurs at $\alpha = 0.448$, the Chern number of the NFB still remains ill-defined up to $\alpha < 0.47$. Beyond this point, the NFB enters a $C = -2$ topological phase, as shown in Fig.~\ref{fig:Figure-S6}(h). With further increase in $\alpha$, another gap-closing transition takes place at $\alpha = 0.556$ [Fig.~\ref{fig:Figure-S6}(i)] at the $\Gamma$ point of the MBZ, marking a first-order topological phase transition from the $C = -2$ to the $C = -1$ phase [Fig.~\ref{fig:Figure-S6}(j)]. 
\begin{figure}
    \centering    \includegraphics[width=0.55\linewidth]{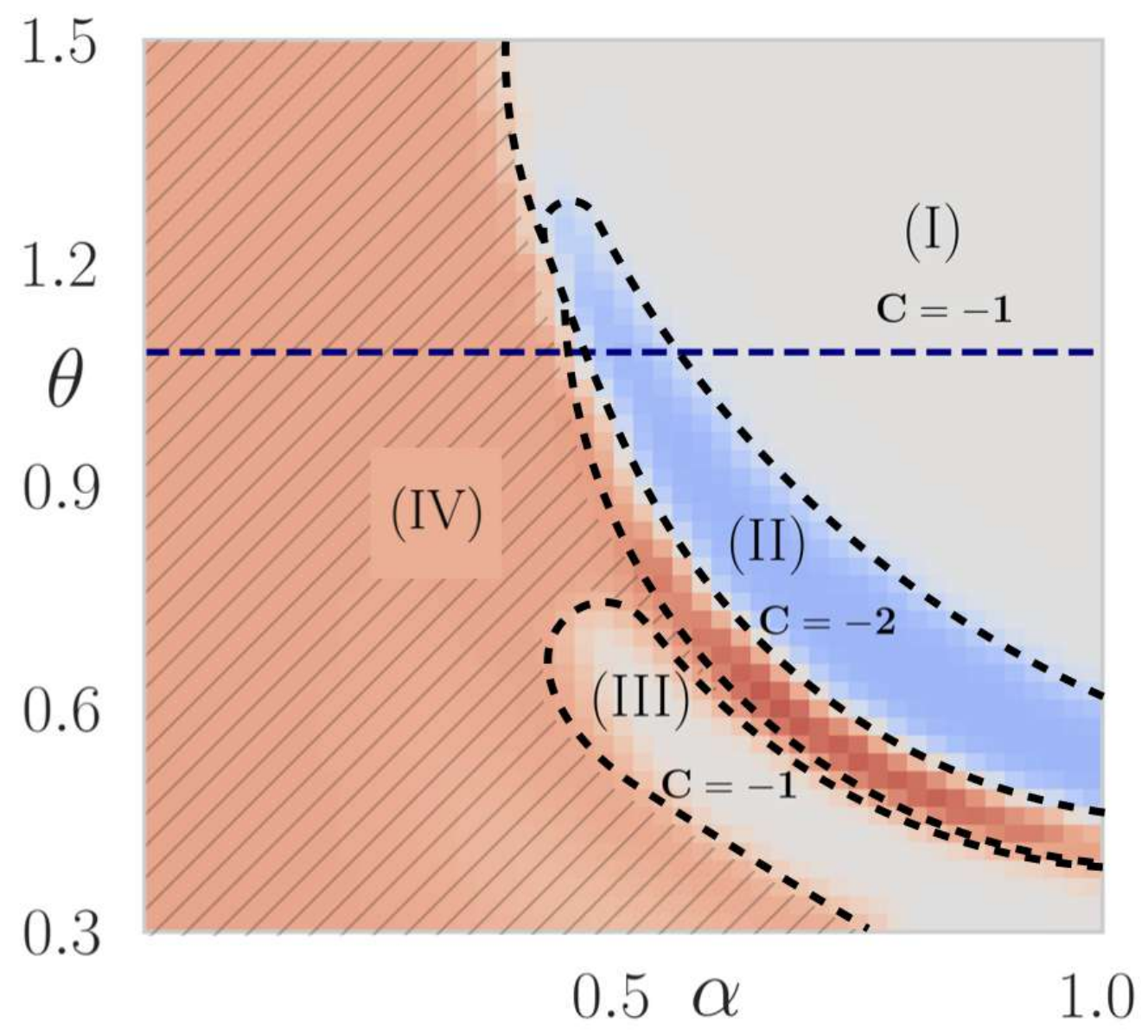}
    \caption{The phase plot of the Chern number for the NFB in the $\alpha-\theta$ plane with finite $w_1$ and $w_3$ ($w_1=w_3=0.332 \mathrm{~meV}$) is shown. The horizontal dotted line corresponds to the magic angle limit as previously shown in the original manuscript. The colors separate the areas with varying Chern numbers which are mentioned in the figure.
    The phase marked (IV) corresponds to an area of negligible ($10^{-4}$ meV) or no band gap, resulting in an essentially undefined topological characterization.
    }
    \label{fig:Figure-S8}
\end{figure}
The system remains in this $C = -1$ phase for higher $\alpha$ values, persisting up to the dice limit ($\alpha = 1$). The corresponding phase plot with $w_1=w_3=0.003~w_2$ is shown in Fig.~\ref{fig:Figure-S8}. With further increasing $w_1$ and $w_3$, the region-IV (corresponding to the undefined topological phase) in the Chern phase diagram [Fig.~\ref{fig:Figure-S8}] shifts towards the right, while region-I (also characterized by $C=-1$) moves towards the left. Consequently, the area of region-II corresponding to the $C=-2$ phase gradually decreases with increasing $w_1$ and $w_3$, and becomes constricted as we double the value of $w_1 = w_3 = 0.006~w_2$. The Region–III associated with the $C=-1$ phase entirely disappears at $w_1=w_3= 0.006~w_2$. Upon further increasing $w_1$ and $w_3$ the $C=-2$ phase becomes even smaller and eventually vanishes, leaving the NFB entirely in the $C=-1$ phase throughout the $\alpha$-$\theta$ plane.
\end{document}